\documentclass{article}

\usepackage{graphicx}%
\usepackage{multirow}%
\usepackage{amsmath,amssymb,amsfonts}%
\usepackage{amsthm}%
\usepackage{mathrsfs}%
\usepackage[title]{appendix}%
\usepackage{xcolor}%
\usepackage{textcomp}%
\usepackage{manyfoot}%
\usepackage{booktabs}%
\usepackage{algorithm}%
\usepackage{algorithmicx}%
\usepackage{algpseudocode}%
\usepackage{listings}%
\usepackage{enumitem}%
\usepackage{geometry}%
\usepackage{todonotes}
\usepackage[square,numbers]{natbib}

\begin{document}

\title{MMIL: A novel algorithm for disease associated cell type discovery}

\author{Erin Craig$^{*1}$  \and  Timothy Keyes$^{*1}$ \and Jolanda Sarno$^{3, 6}$ \and Maxim Zaslavsky$^4$ \and Garry Nolan$^5$\and  Kara Davis$^6$ \and Trevor Hastie$^{1,2}$ \and
 Robert Tibshirani$^{1,2}$}
\date{\normalsize
$^1$Department of Biomedical Data Science, Stanford University\\
$^2$Department of Statistics, Stanford University\\
$^3$M. Tettamanti Research Center, Pediatric Clinic, University of Milano Bicocca\\
$^4$Department of Computer Science, Stanford University \\
$^5$Department of Pathology, Stanford University \\ 
$^6$Department of Pediatrics, Stanford University \\
$^*$These authors contributed equally to this work.}

\maketitle

\abstract{

Single-cell datasets often lack individual cell labels, making it challenging to identify cells associated with disease. To address this, we introduce Mixture Modeling for Multiple Instance Learning (MMIL), an expectation maximization method that enables the training and calibration of cell-level classifiers using patient-level labels. Our approach can be used to train e.g. lasso logistic regression models, gradient boosted trees, and neural networks. When applied to clinically-annotated, primary patient samples in Acute Myeloid Leukemia (AML) and Acute Lymphoblastic Leukemia (ALL), our method accurately identifies cancer cells, generalizes across tissues and treatment timepoints, and selects biologically relevant features. In addition, MMIL is capable of incorporating cell labels into model training when they \textit{are} known, providing a powerful framework for leveraging both labeled and unlabeled data simultaneously.  Mixture Modeling for MIL offers a novel approach for cell classification, with significant potential to advance disease understanding and management, especially in scenarios with unknown gold-standard labels and high dimensionality.

}

\maketitle

\section{Introduction}\label{sec:intro}

In the biomedical field, we often know whether or not a person is sick. However, we do not necessarily know which cells in a sick person’s body--which contains a combination of diseased and healthy cells--play a role in their illness. Identifying disease-associated cell populations is a common goal in translational studies of single-cell datasets, with applications in systems biology, disease monitoring, diagnostics, and the discovery of targetable cell populations for novel therapeutics. Nevertheless, classifying cells as sick or healthy can be difficult: obtaining gold standard labels can be prohibitively challenging or expensive, and it is not obvious how to predict labels without labeled training data. This is particularly true in the context of complex diseases for which true disease-associated cell populations remain unidentified, making gold-standard cell annotation impossible.

Without cell labels to train a classifier, researchers often turn to methods from \emph{multiple instance learning} (MIL), which are designed for data consisting of many unlabeled cells from each labeled patient. However, the primary aim of most MIL methods is to predict new patient labels rather than cell labels. Here, we present \textbf{M}ixture Modeling for \textbf{M}ultiple \textbf{I}nstance \textbf{L}earning (MMIL), a method that, armed only with patient labels, uses expectation maximization (EM)~\cite{dempster1977maximum} to train a classifier to label individual cells as disease-associated or non-disease-associated. Our approach alternately estimates cell labels and trains a classifier; at each iteration, our classifier and cell label estimates improve. MMIL is versatile and can be easily implemented as a wrapper around any classifier trained by optimizing the binomial log likelihood; for example, lasso logistic regression models, gradient boosted trees and neural networks. MMIL can also be used to improve the model calibration of any cell-level model --- calibration is important because it encourages the alignment of estimated probabilities with outcomes such that e.g. 90\% of the cells with a $0.9$ probability are truly disease-associated. MMIL's calibration step can be used as a post-processing step for \textit{any} cell-level model, making predictions more reliable and understandable.

We then show that MMIL identifies cancerous cells in leukemia (cancer of the blood and bone marrow). This is an important task: the early and precise identification of malignant cells is pivotal for determining treatment options and monitoring disease progression. Leukemia cells (often called ``leukemic blasts'') are challenging to identify due to phenotypic characteristics resembling the healthy cells from which they originate~\cite{dworzak1999detection}. For this reason, identifying treatment-resistant cells--referred to as ``minimal residual disease'' (MRD) in clinical contexts--is challenging and labor-intensive, even among highly trained clinicians using state-of-the-art blood profiling technologies~\cite{mrd_review2020, chen2017monitoring}. 

We use MMIL to train lasso logistic regression models on data collected from patients with either Acute Myeloid Leukemia (AML) or Acute Lymphoblastic Leukemia (ALL) as well as healthy controls. In both diseases, we demonstrate MMIL's ability to (1) identify cancer cells accurately; (2) generalize to cells sampled from different patients, different timepoints after initiation of treatment, or different tissues than those on which the models were originally trained; and (3) select single-cell features known to be associated with leukemia. Our applications in leukemia diagnosis and treatment monitoring reveal that MMIL has broad implications in the field of single-cell analysis. Because MMIL is a direct, flexible method for identifying disease-associated cells using only patient labels, it has the potential to reveal novel biological insights and support patient monitoring, treatment and diagnosis for poorly understood diseases. 

\section{Results}\label{sec:results}

\subsection{An algorithm to classify cells without complete cell labels}
\label{sec:method}
We wish to train a classifier to predict whether a given cell is healthy or diseased. Typically, we would (1) gather a dataset of cells where we \emph{know} which cells are diseased and which are healthy; and (2) train a model to predict cell disease status (using e.g. logistic regression, random forest, neural network). This workflow relies on knowing which cells are diseased and which are not. In our setting we have no such luxury: we do not know which cells sampled from sick patients are diseased, we only know that cells sampled from healthy people are healthy. MMIL is an algorithm that trains classifiers in this setting.

MMIL is an iterative process that relies on the following intuition. If we had a trained classifier, we could use it to \emph{predict} which cells are healthy and which are not. And if we had a \emph{prediction} of which cells were healthy and which were not, then we could use these predicted labels to train a classifier. MMIL fits a classifier by alternating between these two steps, and is described in Algorithm~\ref{alg:alg1} (and in more detail in Appendix~\ref{sec:likelihoods}). 

\begin{figure}[H]
    \centering
      \includegraphics[width=\linewidth]{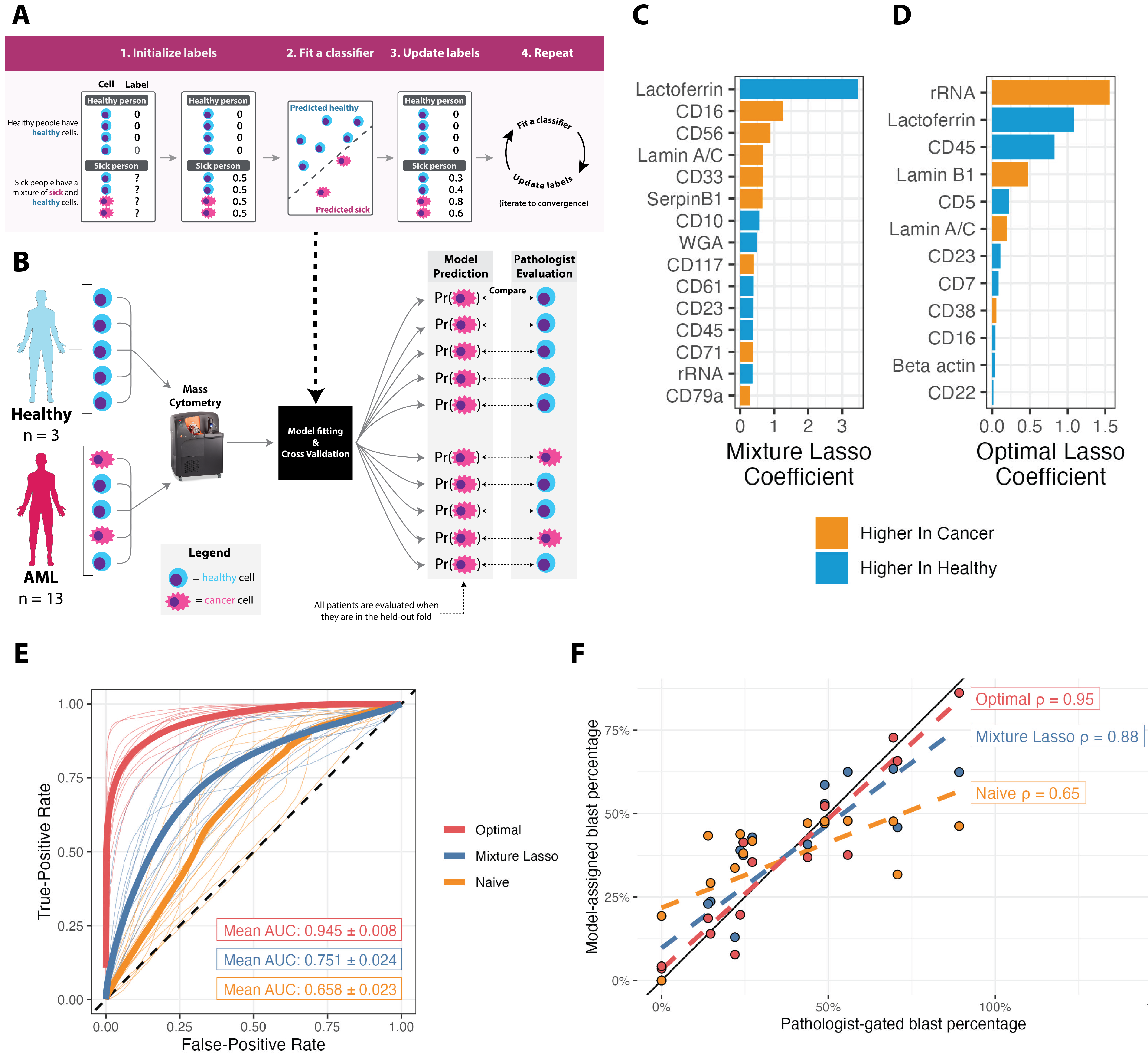}
      \caption{\textbf{Mixture Modeling for Multiple Instance Learning (MMIL) detects cancer cells in Acute Myeloid Leukemia (AML) using patient labels only} \textbf{(A)} Process to train a mixture model for multiple instance learning data. We initialize the sick person's cell labels as $0.5$: in this example, we assume that the prevalence of diseased cells in sick people is $50\%$, and so each cell has a 50/50 chance of being diseased. After training the first classifier, we improve our estimates of the sick person's cell labels. This process is repeated until convergence. \textbf{(B)} Schematic of model training and evaluation on the AML cohort from \cite{tsai2020multiplexed}. \textbf{(C)} Nonzero coefficients for the Mixture Lasso model trained to detect leukemic blasts in AML. \textbf{(D)} Nonzero coefficients for the "Optimal" Lasso model trained to detect leukemic blasts in AML. \textbf{(E)} Receiver-Operator Characteristic (ROC) curves demonstrating individual (thin) and average (thick) performance of Optimal, Mixture, and Naive lasso models trained to detect leukemic blasts in AML. Insets indicate mean area under the ROC curve (AUC) across all patients. \textbf{(F)} Scatterplots representing the relationship between the gold-standard, pathologist-enumerated blast percentage for each patient (X-axis) and the model-assigned blast percentage for each patient for the Optimal (red), Mixture (blue), and Naive (yellow) lasso models. Inset text represents the Pearson correlation coefficients between the values on the X- and Y-axes.  
      }
      \label{fig:algorithm}
\end{figure}

\begin{algorithm}[ht]
\caption{MMIL: Mixture Models for Multiple Instance Learning}
  \label{alg:alg1}
\begin{enumerate}
\item Using domain knowledge, choose a value for $1 - \rho$, the probability that a cell from a sick person is diseased.
\item Create a dataset with:
\begin{itemize}
    \item One copy of each cell from healthy people, assigned the label $y = 0$ (healthy) and weight $w = 1$. 
    \item Two copies of each cell from sick people, one labeled $y = 0$ (healthy) with weight $w = \rho$ and the other $y = 1$ (diseased) with weight $w = 1 - \rho$. 
    \end{itemize}
\emph{The weights reflect uncertainty. We are confident that healthy patients' cells are healthy. But we do not know whether sick patients' cells are diseased or healthy -- the best we can do is assume that each cell is diseased with probability $1 - \rho$.}

\item Alternate:
\begin{enumerate}
    \item Train a classifier using the augmented dataset.
    \item Use the trained classifier to predict whether each sick patient's cell is diseased or healthy. Update their weights using these predictions. (See Methods for details.)
\end{enumerate}
\emph{After training, the classifier improves our guess for whether each sick person's cell is diseased or healthy. Then, the new guesses are used to train a new classifier.}

\item Stop alternating when the model predictions stop changing.
\end{enumerate}
\end{algorithm}

\subsection{Mixture modeling identifies cancer cells in Acute Myeloid Leukemia using only unlabeled cells}
\label{sec:real_data_1}

We now consider a publicly available dataset of 800 thousand cells collected from 13 Acute Myeloid Leukemia (AML) patients and three healthy bone marrow controls \cite{tsai2020multiplexed} (see Methods). Each AML sample is a mixture of cancer cells and healthy cells, and this cohort of patients represents a wide range of leukemic blast counts. Typically, a highly-trained hematopathologist must analyze a blood or bone marrow specimen to diagnose and track the progression of disease over time. Hematopathologists do this by manually enumerating the proportion of leukemic blasts in the sample based on each cell's size, shape, and expression of leukemia-associated markers measured by microscopy or flow cytometry. Particularly in the context of myeloid leukemias, blast enumeration is challenging and error-prone even among expert pathologists \cite{tsai2020multiplexed, hodes_challenging_2019}.

Using MMIL, we sought to automate the identification of cells specific to AML without expert cell-level annotation. Importantly, each cell in the AML dataset has a gold-standard annotation indicating whether it is a leukemic blast as evaluated by a board-certified hematopathologist. These annotations allow us to evaluate MMIL's performance in identifying leukemic blasts compared to expert clinical judgement.

Each cell was analyzed for the presence of 32 surface proteins and 11 intracellular proteins~\cite{tsai2020multiplexed} using mass cytometry (CyTOF), a high-dimensional cytometry platform similar to clinical cytometers commonly used to analyze leukemic tissue specimens in clinical laboratories~\cite{ddpr2018, cytofmethods2021}. However, CyTOF analysis allows a high-dimensional, single-cell characterization of each sample, extending beyond the capabilities of conventional multicolor flow cytometers and making it ideal for deep phenotyping of leukemic cells in a research environment~\cite{coustansmith2002}.

Using this dataset, we trained 3 models: (1) the ``optimal'' model, a lasso logistic regression model trained using gold-standard annotations (usually unknown), (2) the "naive" model, a lasso model trained using patient labels instead of cell labels, and (3) Mixture Lasso, a lasso model trained with MMIL. When training Mixture Lasso, we assumed that $75\%$ of cells in cancer patients are healthy, based on the clinical guideline that patients with at least $25\%$ of blasts in their bone marrow receive a leukemia diagnosis\cite{berg1992leukemiathreshold}. We estimated the performance of each of these models using leave-one-(patient)-out cross-validation (LOOCV). In each fold of our cross-validation procedure, models were trained using all 3 healthy control marrows and 12 of the 13 AML marrows; then, the performance of each model was evaluated by comparing predictions on the held-out patient's cells to the pathologist's gold-standard labels (\textbf{Figure~\ref{fig:algorithm}b}).

At the cell-level, Mixture Lasso achieved a mean area under the receiver-operator characteristic curve (AUROC) of 0.751 across all held-out patients during the cross-validation. By comparison, the naive model achieved worse performance with an AUROC of 0.658, and the optimal model achieved superior performance with an AUROC of 0.945 (\textbf{Figure~\ref{fig:algorithm}e}). At the patient-level, when a probability threshold of 0.5 was used to classify cells as leukemic blasts (see Methods), the blast percentages that Mixture Lasso assigned to each patient in the dataset correlated strongly with pathologist-assigned blast percentages (Pearson $\rho$ = 0.88 compared to the optimal model's $\rho$ = 0.95). The naive model's blast percentage assignments correlated less with the pathologist gold standard (Pearson $\rho$ = 0.65). 

In their original study, Tsai et al. identified two sets of markers that could be used to distinguish between different stages of myeloid development in healthy and AML cell populations. Among these were the cytoskeletal proteins $\beta$ actin, Lamin A/C, and Lamin B1; the granule-associated proteins SerpinB1, VAMP-7, ribosomal RNA (rRNA), Lactoferrin, and lysozyme; CD45; and the cell size marker wheat germ agglutinin lectin (WGA). Two of the most important markers for the pathologists' AML blast enumeration were rRNA (highly expressed in AML blasts due to their rapid growth) and Lactoferrin  (a granule protein only expressed in mature myeloid cells that remains lowly expressed in undifferentiated AML blasts). \textbf{Figures~\ref{fig:algorithm}b-c} show that Lactoferrin, Lamin A/C, CD45, and rRNA were selected by both the Mixture Lasso and optimal lasso models, whereas SerpinB1 and WGA were uniquely selected by Mixture Lasso. However, Lamin B1 and $\beta$ actin were uniquely selected by the optimal lasso model, and only the optimal model selected rRNA with a positive coefficient. Thus, Mixture Lasso closely follows, but does not entirely reproduce the pathologist's decision-making process, suggesting an ability to augment (not just replicate) existing clinical decision processes. 

\begin{figure}[H]
    \centering
      \includegraphics[width=.75 \linewidth]{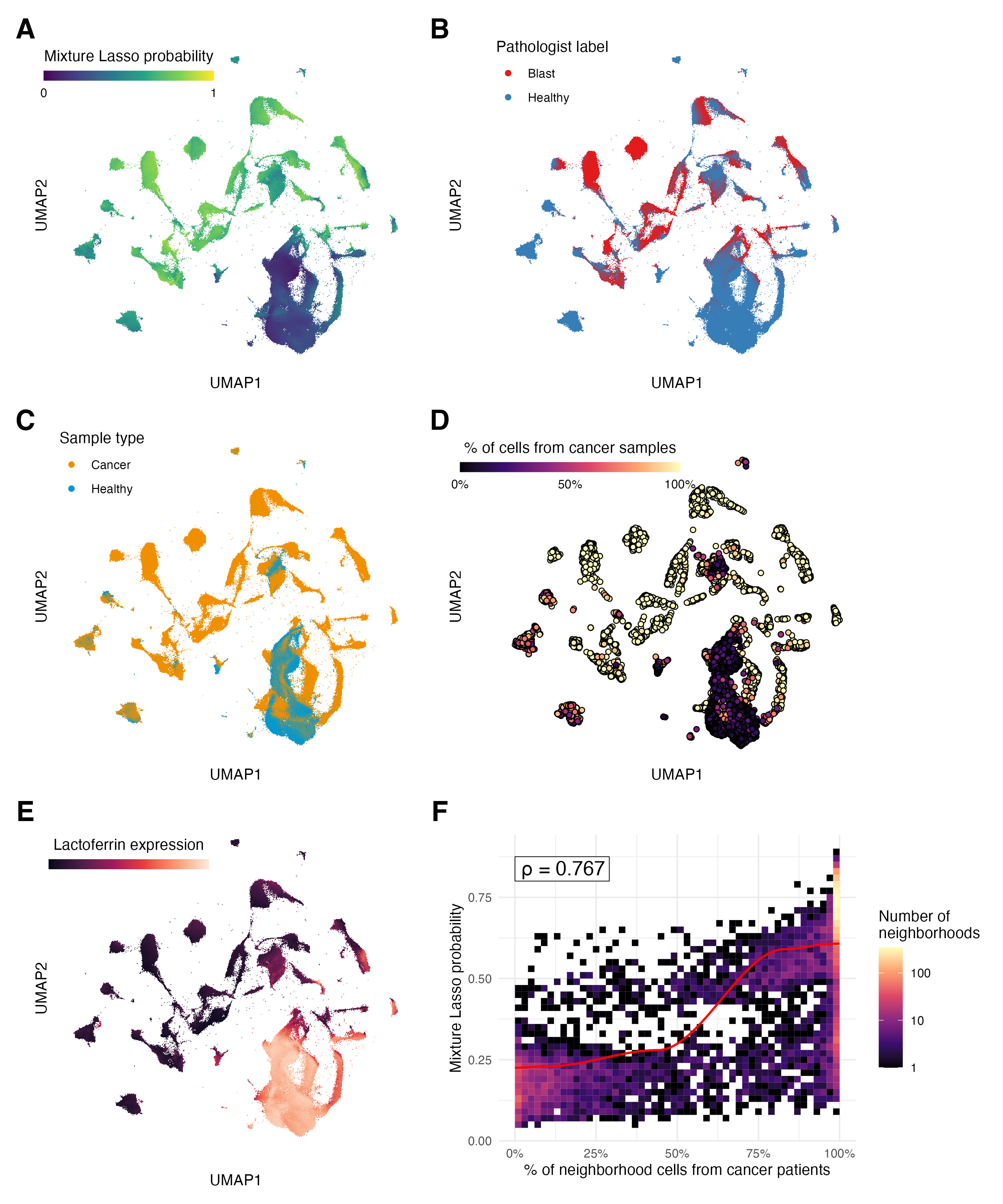}
      \caption{\textbf{MMIL identifies regions of high-dimensional phenotype space occupied by cells from AML patients, but not by cells from healthy controls.} \textbf{(A)} A scatterplot of uniform manifold approximation and projection (UMAP) coordinates colored by Mixture Lasso probabilities. Cells with probability scores of 0 have a very small chance of being AML-associated (i.e. leukemic blasts), whereas cells with probability scores close to 1 have a high chance of being AML-associated (leukemic blasts). \textbf{(B)} UMAP plot as in (A), but with cells annotated as leukemic blasts by a pathologist in red and cells annotated as healthy (i.e. non-leukemic blasts) by a pathologist in blue. Note the general agreement of probabilities in (A) to red regions in (B). \textbf{(C)} UMAP plot as in (A), but with cells collected from cancer patients shown in orange and cells collected from healthy controls in blue. Note that regions with overlapping orange and blue cells are assigned low Mixture Lasso probabilities in (A).  \textbf{(D)} A UMAP plot of local "phenotypic neighborhoods" spaced throughout the high-dimensional point cloud selected by density-dependent downsampling (Methods). Neighborhoods are colored based on the proportion of cells that they contain that come from cancer patients. Note that neighborhoods exclusively comprised of cells from cancer patients are assigned high Mixture Lasso probabilities in (A). \textbf{(E)} A UMAP plot as in (A), but with cells colored by their expression of Lactoferrin, the marker with the largest coefficient in the Mixture Lasso model. See also Supplementary Figure~\ref{fig:supplemental_figure_2}. \textbf{(F)} Count heatmap of 2-dimensional bins demonstrating the correlation between the average Mixture Lasso probability in a phenotypic neighborhood (Y-axis) and the proportion of cells from cancer patients it contains (X-axis). Bins are colored by the density of neighborhoods in that region of the plot, and the red line represents the locally-weighted moving average across the x-axis. Inset text indicates the Pearson correlation between the values on the X- and Y-axes. \textbf{Note:} In panels A-E, UMAP coordinates were calculated using all protein markers. 
      }
      \label{fig:umaps}
\end{figure}

To visualize Mixture Lasso's predictions, we constructed several uniform manifold approximation and projection (UMAP) \cite{mcinnes2018umap} plots using all cells and markers in the dataset (\textbf{Figure~\ref{fig:umaps}a-e}). We find that cells close to one another in UMAP space tend to have similar Mixture Lasso probabilities, indicating that Mixture Lasso tends to identify phenotypically coherent cell populations. Likewise, regions of UMAP space with high Mixture Lasso probabilities tend to correspond to regions containing large numbers of pathologist-identified AML blasts. However, Mixture Lasso does not simply assign high probabilities to all cells collected from cancer patients; rather, it assigns high probabilities to cells in areas of high-dimensional phenotype space preferentially occupied by cells from cancer patients, but not from healthy patients. Intuitively, this makes sense: leukemic blasts should map to regions of phenotype space with little to no overlap with phenotypes present in healthy samples. This observation is quantified in (\textbf{Figures~\ref{fig:umaps}d,f}).

Protein-specific UMAP plots further indicate that Mixture Lasso assigns high disease-association probabilities to cells with unique marker expression patterns--generally, those low in Lactoferrin and high in SerpinB1, CD16, and/or CD56. Furthermore, while rRNA was not explicitly selected as a disease-associated feature in the Mixture Lasso model, cells with high probabilities also tend to express high levels of rRNA (\textbf{Figure~\ref{fig:supplemental_figure_1}} and \textbf{Figure~\ref{fig:supplemental_figure_2}}).


\subsection{Mixture modeling identifies cancer cells in Acute Myeloid Leukemia using both labeled and unlabeled cells}
\label{sec:real_data_2}

Next, we sought to evaluate MMIL's performance in the setting where both labeled and unlabeled cells are present--e.g. a scenario in which most leukemic patients in a dataset do not have cell labels, but a small number of them \textit{have} been annotated. We hypothesized that a model capable of leveraging both the expert labels of a clinician and a larger corpus of unlabeled data would outperform models trained using only labeled or only unlabeled data \cite{yakimovich2021labels}. 

However, using expert labels to train a model on a difficult classification task can be risky: clinical misidentification of leukemic blasts is a known challenge in hematopathology, with a high degree of documented interobserver variability and methodological disagreement (\textbf{Supplemental Figure~\ref{fig:supplemental_figure_3}}) \cite{hodes_challenging_2019}. This is particularly concerning for training machine learning models, as the degradation of model performance due to label error is a widely characterized problem~\cite{northcutt2020pervasive, bernhardt2022active}. Thus, we also hypothesized that models trained on both labeled and unlabeled data would be more robust to label error than models trained on labeled data alone, as they would be less susceptible to over-fitting on the imperfect labels. 

\begin{figure}[H]
    \centering
      \includegraphics[width=0.5\linewidth]{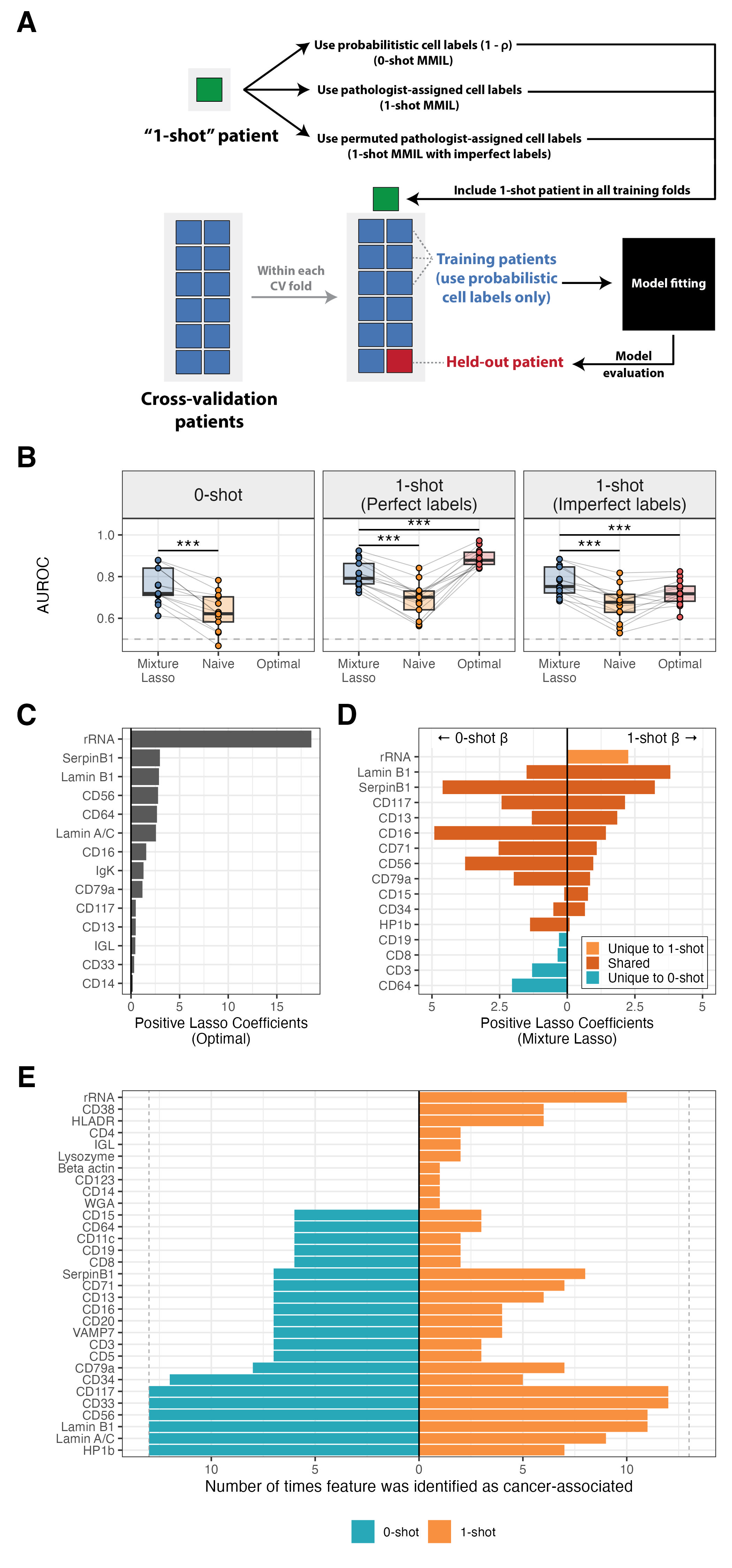}
      \caption{\textbf{MMIL can train on labeled and unlabeled data simultaneously to incorporate expert knowledge while remaining robust to imperfect labeling.} \textbf{(A)} Schematic of semi-supervised ``0-shot'' and ``1-shot'' MMIL experiments (also see Methods). \textbf{(B)} Boxplots indicating average area under the receiver-operator characteristic curve (AUROC) for Mixture Lasso (blue), Naive (orange), and Optimal models across 0-shot (left), 1-shot (with perfect labels; middle), and 1-shot (with imperfect labels; right) training procedures. ``***'' represents statistical significance at the level of $p<0.0001$ using a paired Student's t-test with Benjamini-Hochberg correction for multiple comparisons. \textbf{(C)} Positive coefficients for an Optimal Lasso model fit on a single patient (\textit{AML-5An}). \textbf{(D)} Positive Mixture Lasso coefficients after 0-shot (left) and 1-shot (right) learning. Note that rRNA, the feature with the largest Optimal Lasso coefficient in (C) and \ref{fig:algorithm}d, was selected with a positive coefficient only after 1-shot learning. \textbf{(E)} Two-sided barplot indicating how many times a feature was included in the Mixture Lasso model with a positive coefficient after 0-shot (left, blue) and 1-shot (right, orange) training. Dashed gray lines indicate the maximum number of times a feature could have been included (13, the total number of 1-shot experiments).  
      }
      \label{fig:figure_3}
\end{figure}

To test these hypotheses, we developed a semi-supervised version of MMIL (``1-shot MMIL'') using the training procedure illustrated in (\textbf{Figure~\ref{fig:figure_3}a}). Briefly, 1-shot MMIL is trained identically to MMIL with one change: during model training, a single AML patient (termed the ``1-shot patient'') is chosen and their gold-standard cell labels are used for supervision instead of probabilistic labels. Like the cells from healthy patients, the 1-shot patient's cell labels remain fixed throughout all iterations of the EM. Otherwise, model training and LOOCV are carried out as before. 

To benchmark 1-shot MMIL's performance, we compare it to 0-shot MMIL (the usual MMIL), the 0-shot naive model (a naive model trained as described before), the 1-shot naive model (a model in which the 1-shot patient's gold-standard cell labels are used during model training, but all others use their inherited patient labels), and the 1-shot optimal model (a model trained on only the 1-shot patient using their gold-standard cell labels). Furthermore, to benchmark 1-shot MMIL's robustness to imperfect labeling, we again fit each 1-shot model after randomly permuting 25\% of the 1-shot patient's cell labels--a proportion chosen to match the variability observed between pathologists when enumerating leukemic blasts in AML (\textbf{Figures~\ref{fig:supplemental_figure_3}} and Methods). All models were lasso-regularized logistic regression classifiers.

We performed 13 1-shot experiments: one for each AML patient in the dataset to serve as the 1-shot patient. The results of these experiments are summarized in \textbf{Figure~\ref{fig:figure_3}b-e}. In the ``0-shot'' case, we once again observed that Mixture Lasso outperformed the naive model~(\textbf{Figure~\ref{fig:figure_3}b, left}). In the 1-shot case, we found that Mixture Lasso's performance improved substantially compared to 0-shot Mixture Lasso (average AUROC increase of 0.058) while maintaining its superior performance to the 1-shot naive model~(\textbf{Figure~\ref{fig:figure_3}b, middle}). We also found that training an optimal model using only the 1-shot patient outperforms 1-shot Mixture Lasso; however, after training labels were permuted, 1-shot Mixture Lasso outperforms both the naive model and the optimal model, whose performance degrades substantially after training on the imperfect labels (\textbf{Figure~\ref{fig:figure_3}b, right}). Together, these results suggest that MMIL is capable of improving its performance using even a small number of gold-standard labels, while remaining more robust to noisy labeling than alternative models. 

Based on these results, we wondered how the inclusion of the 1-shot patient during model training affected the feature sets selected by Mixture Lasso. To interrogate this, we examined the 1-shot experiment with the largest average AUROC improvement between the 0-shot and 1-shot models across all patients (patient \textit{AML-5An} was the 1-shot patient). In this experiment, rRNA was selected as the most cancer-associated feature by the optimal model (\textbf{Figure~\ref{fig:figure_3}c}). Interestingly, rRNA was not identified as a cancer-associated marker by the 0-shot Mixture Lasso model, but \textit{was} identified by the 1-shot model (\textbf{Figure~\ref{fig:figure_3}d}). In addition, several markers (CD64, CD3, CD8, and CD19) that were included in the 0-shot Mixture Lasso--but that were not used in Tsai et al.'s manual gating of leukemic blasts--were removed in the 1-shot model. This held as a general trend across \textit{all} 1-shot experiments, across which rRNA was never selected as a cancer-associated marker by 0-shot Mixture Lasso, but was selected as a cancer-associated marker in 10/13 of the corresponding 1-shot Mixture Lasso. At the same time, markers CD3, CD5, CD15, and CD11c were often selected by 0-shot Mixture Lasso but are not canonically considered markers of AML blasts, and these were selected less frequently by 1-shot Mixture Lasso. Taken together, these results demonstrate that MMIL can adapt its feature selection process via semi-supervision, leveraging even a small number of expert-labels to prioritize clinically relevant markers over non-informative ones.

\subsection{Mixture modeling identifies and tracks cancer cells throughout treatment progression in Acute Lymphoblastic Leukemia}
\label{sec:real_data_3}

Finally, we applied MMIL to a dataset of 1.1 million blood and bone marrow cells collected from three Acute Lymphoblastic Leukemia (ALL) patients and three healthy controls. This dataset contains samples collected from multiple tissues (blood and bone marrow) and multiple timepoints throughout treatment (diagnosis, Day 8 post-treatment initiation, and Day 15 post-treatment initiation). Using this dataset, we wanted to evaluate the ability of an MMIL model trained at diagnosis to accurately identify and track residual leukemic cells during treatment progression.

The identification of residual leukemic cells throughout treatment is crucial for assessing a patient's risk of relapse and therefore for clinical decision making. Thus, MMIL's ability to maintain strong performance despite distributional shift--i.e. a shift in the distribution of leukemic blast phenotypes over time due to treatment effects or changes in disease state--would be paramount for its application to disease monitoring in cancer.

To evaluate MMIL's ability to detect residual leukemic cells over the course of treatment, we analyzed all cells in the ALL cohort for the presence of 29 proteins as previously described using CyTOF (see Methods). Optimal, naive, and Mixture Lasso models were trained using only bone marrow cells from the diagnostic timepoint, and model performance was once again evaluated using leave-one-patient-out cross validation. Next, we applied our diagnostic models to paired blood samples from the diagnostic timepoint to evaluate each model's ability to generalize to a new tissue context that is more easily monitored clinically. In addition, we applied each model to paired blood samples taken at days 8 and 15 post-treatment initiation to evaluate its ability to track a patient’s disease burden over time. 

\begin{figure}[H]
    \centering
      \includegraphics[width=\linewidth]{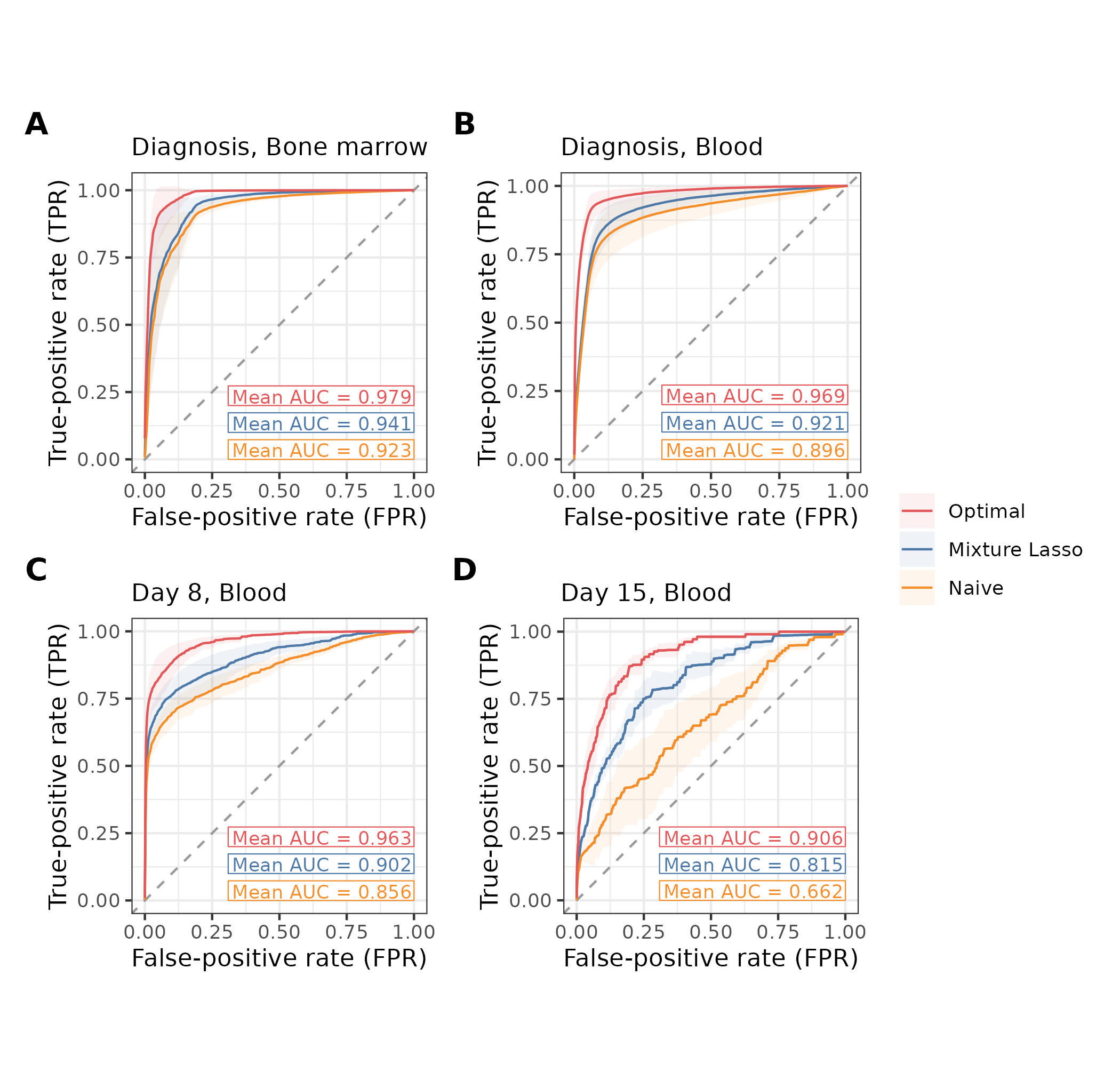}
      \caption{Mixture Lasso identifies leukemia cells despite being trained without any cell labels. We compare the performance of Mixture Lasso to the optimal model (trained using gold standard labels that are typically unavailable) and the naive model (trained using patient labels in place of cell labels). Mixture Lasso generalizes better than the naive approach from bone marrow to blood samples and across time, evidenced by its performance on blood samples collected at (B) diagnosis, (C) day 8 post-treatment initiation, and (D) day 15 post-treatment initiation.}
      \label{fig:real_patients_over_time}
\end{figure}

Trained without access to gold standard labels, Mixture Lasso demonstrates robust performance and generalizability, with an AUROC of 0.941 at diagnosis and 0.815 at day 15 post-treatment initiation. The naive model begins with AUROC of 0.923 at diagnosis, but this degrades across time to 0.662 at Day 15 (Figure~\ref{fig:real_patients_over_time}). This decline in performance aligns with clinical challenges encountered over the course of a patient's treatment, including the diminishing proportion of leukemic blasts over time (as low as 0.1-5\% of all cells) and phenotypic plasticity in resistant cells post-therapy~\cite{sarno2023}. Notably, the naive model's brittle performance in response to these shifts highlights its susceptibility to false positives. Mixture Lasso, however, retains strong performance even as the percentage of leukemic cells becomes low, suggesting its potential for robust disease monitoring and minimal residual disease detection in leukemia.

As before, we analyzed Mixture Lasso's coefficients to examine which protein markers best distinguish between healthy and leukemic cells. Among the markers with positive (cancer-associated) coefficients identified by Mixture Lasso (Supplemental Figure~\ref{fig:supplemental_figure_5}) were CD10, CD19, PAX5, and CD34—each of which has been previously described as aberrantly overexpressed in ALL~\cite{euroflow2012}. For example, CD10, or ``Common Acute Lymphoblastic Leukemia Antigen''(CALLA), is a surface protein expressed in primitive, undifferentiated B cell precursors as well as in the majority of ALL cells \cite{pui_leukemia2003}. Mixture Lasso also identified CD58, a well-known marker of clinical MRD \cite{veltroni_mrd2003}. Conversely, all the proteins with negative coefficients in the Mixture Lasso model are phenotypic markers of mature B cells --- such as surface immunoglobulin (IgMs), intracellular immunoglobulin M (IgMi), CD20, and the lambda chain of the B cell receptor (Supplemental Figure~\ref{fig:supplemental_figure_5}). Each of these proteins are expressed primarily during stages of B cell development that are not reached by leukemic cells, whose differentiation is blocked at earlier stages of B-cell development ~\cite{ddpr2018}.

\section{Discussion}\label{sec:discussion}
Biomedical studies of single-cell data often involve analyzing many unlabeled cells from each patient in order to identify disease-associated cells. Finding such cells has broad-reaching applications in biomedical science, such as aiding our comprehension of disease-contributing cell populations and improving disease diagnosis and monitoring. Mixture Modeling for MIL (MMIL) is a method to train a cell classifier using patient labels only. Our approach has several attractive features. First, it adapts to different choices of classification algorithms: it is straightforward to implement as a wrapper that repeatedly calls established classification software that optimizes the binomial log likelihood. This makes it easy to implement our method within a pre-existing machine learning pipeline. Second, in the past it has been unclear how to calibrate models without labeled data; because our approach can estimate cell labels, we can use it to perform Platt scaling, a standard calibration method. This is useful because interpretable probability estimates empower straighforward downstream clinical decision-making. Finally, MMIL also supports partially labeled datasets: if a small number of cells have known positive labels, they can be included during model training to guide the classifier. This means that MMIL is able to learn from both labeled and unlabeled data simultaneously, allowing it to leverage existing knowledge while remaining robust to noisy or imperfect gold-standard labels. 

Applied to a dataset of cells from patients with and without leukemia, MMIL performed and generalized better than the naive approach of training a classifier using patient labels in place of cell labels. In the context of detecting leukemic blasts in AML, we found that MMIL offered substantial improvement over the naive model. Furthermore, we observed that including a single pathologist-annotated AML sample in MMIL's training set was sufficient not only to improve its performance, but also for it to incorporate expert knowledge into its own decision making. We also showed that MMIL is more robust to noisy labeling than traditional supervised learning methods, allowing for robust performance in the classically difficult task of leukemic blast identification. 

In the context of ALL, we additionally found that MMIL offered modest improvement over the naive model when applied to patient samples from the same timepoint (diagnosis) as the samples on which the models were trained (Figure~\ref{fig:real_patients_over_time} A-B). However, the naive model’s performance degraded substantially when applied to samples collected at later timepoints, likely due to phenotypic shifts occuring in leukemic blasts throughout chemotherapy \cite{cog2005}. By contrast, MMIL’s performance was more robust at later timepoints—particularly at day 15 post-treatment initiation--with a mean AUC of 0.815 compared to the naive model’s AUC of 0.662 (Figure~\ref{fig:real_patients_over_time} C-D). MMIL's robust performance across multiple treatment timepoints highlights its ability to reliably detect biomarkers for disease at the single-cell level.

Only two parameters must be set to train a Mixture Model, aside from any hyperparameters for the chosen classifier. We assume that $\zeta$, the proportion of sick people, and $\rho$, the proportion of healthy cells in sick people, are known. Previous work has shown that these parameters are not, in general, identifiable. Where possible, these parameters should be estimated from other datasets and/or clinical knowledge, and a sensitivity analysis should be performed to evaluate how results are affected. We also assume that healthy patients have no disease-associated cells. When healthy patients may also have a mixture of cells, we instead recommend a clustering method such as MDA2~\cite{hastie2009elements}.

Mixture Modeling for MIL is a flexible, easy to implement method to train a cell-level classifier using patient labels, and has potential as a useful tool for biological hypothesis generation, diagnostics, disease monitoring, and biomarker discovery. In this work, we illustrated our method using two distinct biomedical datasets that were carefully hand-labeled by experts. In the future, however, we expect our method to be most impactful in situations when gold-standard labels are unknown; for example, to identify cells associated with adverse patient outcomes in cancer or to discover specific populations of immune cells underlying autoimmune disease.

\section{Methods}\label{sec:methods}

\subsection{Related work}\label{sec:rel_work}
In multiple instance learning (MIL), data consists of labeled bags, each with many unlabeled instances. Negative bags have only negative instances, and positive bags have at least one positive instance. In our case, the ``bags'' are people, each with many unlabeled cells; healthy people have only healthy cells (negative instances), and sick people have at least one diseased cell (positive instance). MIL methods can be broadly categorized into three paradigms~\cite{amores2013multiple}: instance space, bag space, and embedded space. Bag and embedded space methods aim to label bags; the instance space paradigm, like ours, aims to label instances, and so is our primary focus. For example, multiple instance SVM (mi-SVM)~\cite{andrews2002support} performs SVM with the goal of assigning all negative instances to one halfspace and at least one instance from every positive bag to the other. This method trains an SVM iteratively: first, instances are assigned labels and an SVM is fit. Then, using the fitted SVM, instances are relabeled and a new model is trained. This is repeated until the labels stabilize. Another related approach is multiple instance logistic regression with a lasso penalty (MILR)~\cite{chen2017milr}, which treats the instance labels as missing data, and uses expectation maximization to optimize a lasso-penalized binomial log likelihood. This is the same general approach as ours. However, in the expectation step, MILR uses the joint probability of all instances in the same bag; as a result, MILR tends to be biased when the number of instances in each bag becomes large. Our approach, Mixture Modeling for MIL, instead treats instances as the primary objects of study, and does not use their bag-level grouping, thereby avoiding issues related to computing joint probabilities of many instances. This is particularly important in medical settings where we may sample tens or hundreds of thousands of cells per person. Finally, there are many neural network architectures designed for bag space multiple instance learning~\cite{gadermayr2022multiple} that can be used to label cells. One common method is to apply attention~\cite{bahdanau2014neural} to simultaneously label patients and identify the cells responsible for the predicted bag label. 

Our setting is also analogous to the \emph{presence only} or \emph{positive unlabeled} settings, where some data points have known positive labels, and the remaining labels are unknown. We may instead refer to our data as \emph{negative only} data: cells from healthy people are healthy, and therefore have known \emph{negative} labels, and cells from sick people are unlabeled and may have positive or negative labels. In presence only data, unlabeled instances are randomly sampled from the population, and
\begin{equation}
P(y_i = 1 \mid x_i, \text{$i$ is unlabeled}) = P(y_i = 1 \mid x_i).
\end{equation}
In our setting, though, unlabeled data have a higher-than-baseline probability of belonging to the positive class. The knowledge that a cell is from a sick person raises the probability that it is diseased:
\begin{equation}
    P(y_i = 1 \mid x_i, \text{$i$ is unlabeled}) = P(y_i = 1 \mid x_i, \text{$i$ is from a sick person}) > P(y_i = 1 \mid x_i).
\end{equation}
Ward et al.~\cite{ward2009presence} present a method to train a classifier with presence only data using EM. Mixture modeling for MIL is a modification of their approach to accommodate this structural difference in our data.

\emph{Mixture discriminant analysis 2} (MDA2)~\cite{hastie2009elements}, a clustering method, is related to the problem at hand. MDA2 considers data from $K$ classes. Each class is a mixture of Gaussian distributions, the centers of which are shared across all $K$ classes. Here, we have $K = 2$ classes (healthy and sick people). The healthy class is composed of just one distribution: the distribution of healthy cells. Because sick people have both diseased and healthy cells, the sick class is a mixture of two distributions: the healthy and diseased cell distributions. While MDA2 is a natural fit for our setting, it is a generative method, and it assumes the features $X$ are normally distributed in order to model the joint likelihood $P(X, \text{cell labels  $y$})$. This is not always appropriate, and we may instead prefer a discriminative method that models the conditional likelihood $P(y \mid X)$. We may also prefer to use classifiers that identify important features (like lasso and ridge regression), or relationships between features (like gradient boosted trees and neural networks).

Alternately, a simple but problematic approach is to train cell-level classifier using patient labels in lieu of cell labels~\cite{ray2005supervised, carbonneau2018multiple}. In our comparison, we term this the ``naive'' approach. This may work when most cells sampled from sick people are diseased, as the inherited label is correct most of the time. When sick people have a smaller fraction of diseased cells, classifier performance will suffer because of high label noise. Ward et al.~\cite{ward2009presence} further show that using this approach with logistic regression causes coefficients to be biased toward 0, thereby hindering discovery of biologically important features. To directly address label noise, approaches from the denoising literature (e.g.~\cite{muhlenbach2004identifying, natarajan2013learning, song2022learning}) may be useful. Still, our method leverages the knowledge that cells from healthy people have known labels, which is not assumed by more general methods. 

\subsection{Simulation reveals that calibration is possible without ground truth labels}\label{sec:sim_data}
We wish to simulate a dataset with size $n = 1000$ and $p = 100$ such that $\rho = \zeta = 0.5$, and the relationship between $x_i$ and $y_i$ is given by ${\text{logit } P(y_i \mid x_i) = \beta^T x_i}$. In each simulation, the first $10$ coefficients of $\beta$ are drawn from a normal distribution with mean and standard deviation 1, and the remainder are 0. Then, $x_i \in \mathbb{R}^{100}$ is sampled from a standard multivariate normal distribution with identity covariance; $y_i$ is from a binomial distribution defined by $P(y_i \mid x_i)$. Instances with $y_i = 1$ (diseased cells) necessarily have $z_i = 1$ (sampled from sick patients), and instances with $y_i = 0$ (healthy cells) have $z_i = 1$ with probability $\frac{\rho \zeta}{\rho  \zeta + (1-\zeta)}$. We repeat this process until we have $1000$ instances ($500$ train and $500$ test) satisfying the requirement $\rho = \zeta = 0.5$. 

Then, as before, we trained the optimal model, the naive model, and Mixture Lasso. In addition, we also trained a multiple instance SVM model (mi-SVM, described above). We are interested in both prediction and inference, so we compare the models in terms of AUROC, AUPRC, expected calibration error, and the sum of absolute differences between $\hat{\beta}$ and $\beta$. For the optimal model, calibration is done with Platt scaling using the true labels $y$; this represents the best performance we could reasonably expect. The naive model is calibrated using the incorrect labels $z$, to stay consistent with the fact that it was trained using the incorrect labels. The Mixture Lasso and mi-SVM models are calibrated in two steps:  (1) obtain held-out predictions for each point in the training set (or use a separate held out dataset) and (2) fit a Mixture Lasso model using only the held out probabilities to predict the cell labels $y$. This results in a logistic regression model that adjusts the predicted probabilities to be better calibrated. This simulation is repeated $1000$ times. For all objectives, the optimal model (trained using the true labels) performs best. When the true labels are unavailable, Mixture Lasso outperforms the naive approach (Table~\ref{tab:sim_performance}), and performing Platt scaling using Mixture Logistic Regression greatly improves calibration.

\newcommand*{\thead}[1]{\multicolumn{1}{p{1.5cm}}{\centering #1}}
\begin{table}[h]
    \centering
    \begin{tabular}{r|c c c c c}
         \thead{} & \thead{AUROC} & AUPRC & $||\hat{\beta} - \beta||_1$ & uncalibrated ECE & calibrated ECE \\
         \toprule
       Optimal  & $0.94 \pm 0.02$ & $0.86 \pm 0.05$ & $6.71 \pm 1.58$ & $0.37 \pm 0.02$ & $0.04 \pm 0.01$  \\
       \hline
       Mixture Lasso & $0.90 \pm 0.03$ & $0.80 \pm 0.07$ & $9.64 \pm 2.24$ & $0.11 \pm 0.02$ & $0.06 \pm 0.02$\\
       \hline
       Naive  &  $0.88 \pm 0.04$ & $0.74 \pm 0.07$ & $11.08 \pm 2.43$ & $0.38 \pm 0.02$ & $0.25 \pm 0.03$ \\
       \hline       
       mi-SVM~\cite{misvm_R}  & $0.71 \pm 0.05$ & $0.47 \pm 0.07$ & --- & --- & $0.17 \pm 0.05$ \\
       \hline
    \end{tabular}
    \caption{Comparison of three Lasso logistic regression modeling approaches in 1000 simulations. Optimal was trained with the true labels $y$ (unavailable in realistic settings) and naive with the observed labels $z$. The final two columns show the expected calibration error (ECE), before and after calibration; a lower ECE is preferred. Platt scaling with Mixture Logistic Regression improves calibration for Mixture Lasso, and makes calibration possible for mi-SVM. The mean and one standard deviation of each distribution is shown.}
    \label{tab:sim_performance}
\end{table}

\subsection{Algorithm details}
\begin{algorithm}
\caption{Train a mixture model for multiple instance learning data}
\label{alg:cap}
\begin{description}
    \item \textbf{Input}: Covariates $X \in \mathbb{R}^{n \times p}$,  binary sampling labels $z \in \mathbb{R}^n$, $n_1$ of which are positive.
       \item\textbf{Assumptions}: The proportion of healthy cells in sick people is  $P(y = 0 \mid z = 1) = \rho$. \\
    \phantom{Assumption}The proportion of cells from sick people in the prediction population is ${P(z = 1) = \zeta}$.\vspace{.05 in}
    \hrule
    \item 1. Create a label vector $y^{(0)}$, set to $0$ for cells from healthy people and $1-\rho$ for cells from sick people. 
    \item 2. Iterate to convergence. At the $i^{\text{th}}$ step:
    \begin{enumerate}[label=(\alph*)]
        \item \emph{Maximization step:} Fit a model $\eta^{\star (i)}(x)$ using $X$ and $y^{(i-1)}$. Account for biased sampling: make the case-control intercept adjustment to obtain 
        \begin{equation}
        \eta^{(i)}(x) = \eta^{\star (i)}(x) + \log\left(\frac{(1-\rho) n_1}{n - (1-\rho)n_1}\right) - \log\left(\frac{(1-\rho) \zeta}{1 - (1-\rho) \zeta} \right).
        \end{equation}
        \item \emph{Expectation step:} Use $\eta^{(i)}(x)$ to define $y^{(i)}$. Cells from healthy people have known label $y^{(i)} = 0$. Cell $j$ from a sick person has the label $y_j^{(i)}$ such that: \begin{equation}
            \text{logit } y^{(i)}_j = \eta^{(i)}(x_j) - \log\left(\frac{\rho \zeta}{1 - (1-\rho) \zeta}\right).
        \end{equation} 
    \end{enumerate}
    \hrule
    \item \textbf{Output}: A fitted model $\hat{\eta}(x)$ to predict labels for new cells.
\end{description}
\end{algorithm}

To choose hyperparameters (e.g. the regularization strength in lasso logistic regression), we recommend using cross validation. For each desired choice of hyperparameter, train a model using cells from a subset of the patients, and record the value of the observed data log likelihood (in Section~\ref{sec:likelihoods}) on the cells from held out patients. After repeating this across all desired hyperparameter values and all folds, choose the hyperparameters that maximize the cross validated log likelihood.

Importantly, we have assumed that $\rho$ and $\zeta$, the proportion of healthy cells in sick patients and the proportion of cells sampled from sick patients, are known. We make these assumptions because $\rho$ and $\zeta$ are not, in general, identifiable. Ward et al.~\cite{ward2009presence} show that they are identifiable only if we make strong assumptions about the form of $\eta(x)$; even when it they are identifiable, their estimates have high variance. This argument is reaffirmed and expanded upon by Hastie and Fithian~\cite{hastie2013inference}.

For large datasets, the EM procedure may be slow and it may be more appropriate to optimize the observed log likelihood (Appendix~\ref{sec:likelihoods}) through stochastic gradient descent. For smaller datasets, using EM allows us to easily try many different model choices for $\eta$, and to take advantage of off-the-shelf model fitting software in each loop of the algorithm. 

As we have described it here, our algorithm relies on the use of classification software that can train a model using \emph{soft labels} (values between 0 and 1). In Appendix~\ref{app:no_soft_labels}, we share a modification for classifiers that require \emph{hard labels} (restricted to be 0 or 1).

\subsection{AML dataset analysis}

\subsubsection{Data acquisition}

Normalized, singlet-gated AML data were downloaded from FlowRepository (ID: FR-FCM-Z2E7), and gold-standard patholgist annotations were obtained via correspondence with the authorship team of Tsai et al. (2020). Due to the ambiguity of their diagnosis, the one patient in the cohort with Myelodysplastic Syndrome (MDS) was excluded from analysis. For compatability with the blast enumeration in Tsai et al. (2020) and standard pathology laboratory procedure, only CD45+ events (hematopoietic lineage cells) were analyzed. 

\subsubsection{Data cleaning and preprocessing}

Standard preprocessing steps for mass cytometry data analysis were performed as described previously \cite{cytofmethods2021}. Specifically, ion counts were transformed using the hyperbolic arcsine function with a cofactor of 5, and all markers were scaled to their 99.9th percentile for comparability between markers. Additionally, all cells expressing a marker value over the 99.9th percentile were excluded from analysis in order to remove technical artifacts and outliers. All analyses were performed using the R package tidytof \cite{keyes2023tidytof}.

\subsubsection{Model fitting}

\textbf{Models in Figure 1 and Figure 2}

\textit{Model specification:} MMIL was applied to the AML cohort via the \emph{Mixture Lasso}, a lasso logistic regression model trained using the algorithm described in \ref{sec:method}. We used a value of ($\rho = 0.75$) for all MMIL models, relying on the clinical knowledge that AML patients with more than 25\% blasts in their bone marrow receive the clinical diagnosis of leukemia \cite{berg1992leukemiathreshold}. $\zeta$ was estimated using the training set of each fold (see below). 

In addition to Mixture Lasso, two other models were fit and evaluated: the ``optimal'' model (a lasso logistic regression model trained using the true cell labels, as annotated by a pathologist) and the ``naive'' model (a lasso model trained using inherited patient labels instead of cell labels). For clarity, the cell labels used by Mixture Lasso are referred to as the ``probabilistic labels'' of each cell; the cell labels used by the optimal model are referred to as the ``gold-standard'' or ``pathologist-annotated'' labels of each cell; and the labels used by the naive model are referred to as the ``inherited patient labels'' of each cell. Mixture Lasso, the optimal model, and the naive model are referred to as the 3 ``model classes'' that we evaluated. \newline

\textit{Hyperparameter tuning:} Lasso models have a single hyperparameter: $\lambda$, the penalty term determining the amount of regularization applied to the model's coefficients. For each model class, we tuned over 10 values of lambda, equally spaced on a logarithmic scale between the lowest value ($10^{-5}$) and the highest value (1). The optimal hyperparameter for each model class was determined using cross-validation (see below). The model predictions of the model fit with the optimal penalty parameter were used for reporting in Figure 4's ROC curves and AUROC values. \newline

\textit{Cross-validation and model selection:} Model performance was estimated using leave-one-(patient)-out cross-validation (LOOCV), a schema in which all cells from a single AML patient are held out as a separate test set in each fold of the cross-validation. Specifically, we break the dataset into 13 folds such that each fold includes 12 AML patients and all 3 healthy controls for model training and 1 held-out AML patient for model evaluation. For each fold and model class, we fit the model on the training set and evaluate its cell-level performance on the held-out AML patient according to gold-standard, pathologist-annotated cell labels. The optimal lasso regularization penalty for each model class was chosen by selecting the penalty that optimized the average log-likelihood (for Mixture Lasso) or binomial deviance (for both other models) on the held out sample across all folds. Finally, for model interpretation, we refit each model with the optimal penalty on all 13 AML samples and all 3 healthy bone marrow samples.\newline

\noindent \textbf{0-shot and 1-shot models in Figure 3}: 

In this section, we refer to models trained on a dataset for whom no cell labels are known ``0-shot'' models (the 0 denoting the number of patients who have received expert annotation before model training). Similarly, we refer to models trained on a dataset for whom a single patient's cell labels are known ``1-shot'' models (the 1 again denoting the number of patients who have received expert annotation before model training). We borrow this language from the few-shot machine learning literature, a paradigm that has been scarcely applied to multiple-instance learning problems but that is highly applicable here.

0-shot models are simply the same models described in the ``\textbf{Models in Figure 1 and Figure 2}'' section above. In the 0-shot case, MMIL models are trained using probabilisitic cell labels as described in \textbf{Figure~\ref{fig:algorithm}a}, naive models are trained using inherited patient labels, and optimal models are trained using gold-standard cell labels. Notably, it is not possible to fit a 0-shot optimal modelm, as optimal models require direct cell labels. This is why the left panel of \ref{fig:figure_3}b has a blank space for the optimal model.

By contrast, 1-shot models are trained identically to 0-shot models except for a single change: during model training, one AML patient (termed the ``1-shot patient'') is chosen such that their gold-standard cell labels are used for supervision instead of their probabilistic labels (for Mixture Lasso) or inherited patient labels (for naive models). Similarly to the cells from healthy patients, the 1-shot patient's cell labels remain fixed throughout all iterations of the EM. Otherwise, model training, cross-validation, and model selection are carried out as before. 

Note that a single '1-shot experiment' in the main text refers to the following training procedure: 

\begin{enumerate}
    \item Choose one of the 13 AML patients to designate as the 1-shot patient. Designate all other 12 AML patients as "cross-validation patients."

    \item Fit a 0-shot Mixture Lasso model and a 0-shot naive model using 12-fold cross validation. In each fold of the cross validation, 11 cross-validation patients, the 1-shot patient, and all 3 healthy controls are included in the training set, and 1 cross-validation patient is used as the held-out test set. For Mixture Lasso, the model with the penalty parameter that optimizes the average MMIL log likelihood in the test set across all folds is chosen as the best model and is used for error reporting. For the naive model, the model with the penalty parameter that optimizes the average binomial deviance in the test set across all folds is chosen as the best model and used for error reporting.

    \item Fit a 1-shot Mixture Lasso model and a 1-shot naive model using 12-fold cross validation. In each fold of the cross validation, 11 cross-validation patients, the 1-shot patient, and all 3 healthy controls are included in the training set, and 1 cross-validation patient is used as the held-out test set. Additionally, fit an optimal model using only the 1-shot patient. For Mixture Lasso, the model with the penalty parameter that optimizes the average MMIL log likelihood in the test set across all folds is chosen as the best model and is used for error reporting. For the naive model, the model with the penalty parameter that optimizes the average binomial deviance in the test set across all folds is chosen as the best model and used for error reporting. For the optimal model, only a single patient is available for model training, so the model with the penalty parameter that optimizes the average binomial deviance for that patient is chosen as the best model and used for error reporting. 

    \item Repeat step 3 after randomly permuting 25\% of the 1-shot patient's gold-standard labels. 
\end{enumerate}

Thus, to give each AML patient a turn being the 1-shot patient, steps 1-4 were repeated 13 times. Finally, the AUROC for each left-out patient was averaged across all 13 1-shot experiments, giving an expected AUROC across all possible choices of 1-shot patient. These AUROC values are the reported values for each patient in \ref{fig:figure_3}b. 

\subsubsection{Model evaluation and interpretation}

\textbf{Single-cell model evaluation:}To evaluate each model class's performance at the single-cell level, we calculated the ROC curve and corresponding AUROC for each sample using gold-standard cell labels.\newline

\noindent \textbf{Blast percentages in Figure \ref{fig:algorithm}e}: Mixture Lasso, naive, and optimal probabilities were assigned to each cell using the models refit on all AML patients with the best penalty parameters identified by cross-validation. These probabilities were used to classify each cell in the dataset as either a leukemic blasts or not a leukemic blast with each of the 3 model classes. In the case of the optimal model, cell labels are known, so the probability cutoff was chosen to give the highest cell label classification accuracy. In the case of the naive model, cell labels are not known, so the probability cutoff was chosen to give the highest inherited patient label classification accuracy. Finally, for Mixture Lasso, the cell labels are not known, but because of Mixture Lasso's strong calibration due to our model fitting procedure, a simply probability cutoff of 0.5 was chosen.

Using these binary classifications for each cell, blast percentages were calculated for each patient. To recalibrate the blast percentages assigned by each model with the pathologist-enumerated blast percentages, a linear regression was fit for each model class as a post-processing step. Specifically, a linear regression of the model-assigned blast percentage onto the pathologist-enumerated blast percentage was used to derive the values plotted on the y-axis of \ref{fig:algorithm}e.\newline

\noindent \textbf{UMAP plots:} To analyze how cells assigned high MMIL probabilities arrange in high-dimensional space, we performed dimensionality reduction using uniform manifold approximation and projection (UMAP) using all cells and all markers in the dataset (\ref{fig:umaps}). The plotted MMIL probabilities were taken from the Mixture Lasso model refit on all AML patients using the optimal penalty parameter identified by cross-validation, and UMAP was run using the default parameters of the tidytof R package \cite{mcinnes2018umap, keyes2023tidytof}. \newline

In  Figure \ref{fig:umaps}d, local neighborhoods were constructed using a two-step process. First, density-dependent downsampling (\cite{qiu2011extracting, keyes2023tidytof}) was used to select index cells from the full dataset such that all regions of phenotypic space are represented equally, with each index cell representing the center of a local neighborhood in high-dimensional space. All markers were used to calculated the local densities surrounding each cell in the dataset during density-dependent downsampling. This downsampling process selected 6,849 index cells that were approximately evenly dispersed throughout the high-dimensional point cloud. After index cells were selected, the percentage of cells in each local neighborhood collected from AML patient samples was calculated by finding the 100-nearest neighbors of each index cell in the original dataset and computing the proportion of those neighbors from cancer samples. \newline

\noindent \textbf{0-shot and 1-shot Lasso coefficient analysis:} For model interpretation in Figure  \ref{fig:figure_3}, we examined the non-negative coefficients of the Mixture Lasso model refit on all data using the optimal penalty parameter identified by cross-validation. We focused our analysis on positive (i.e. cancer-associated) lasso coefficients because, generally speaking, disease-associated features are more useful as positive biomarkers that can be used to diagnose or monitor disease. In Figure \ref{fig:figure_3}e, features were counted as disease-associated if they had a lasso coefficient of at least 0.01 in their Mixture Lasso model.

\subsection{ALL dataset analysis}

\subsubsection{Data acquisition}

De-identified bone marrow and peripheral blood primary samples from patients with ALL were obtained under informed consent from Lucile Packard Children’s Hospital and Stanford Hospital from the Pediatric Clinic of University of Milano-Bicocca (San Gerardo Hospital, Monza, Italy). The use of these samples was approved by the Institutional Review Boards at Lucile Packard Children’s Hospital, Stanford Hospital at Stanford University, and from the Pediatric Clinic of University of Milano-Bicocca. Cryopreserved primary bone marrow and peripheral blood samples were thawed rapidly in thawing media (RPMI 1640 supplemented with 10\% fetal bovine serum, 1\% penicillin-streptomycin, and glutamine, 20 U/mL sodium heparin, and 0.025 U/mL Benzonase. Cells were rested for 30 minutes at 37°C and cisplatin viability stained \cite{cisplatin2012}. 

Each sample was analyzed for the expression of 29 proteins using mass cytometry as previously described. \cite{cytofmethods2021}.  Briefly, cells were fixed with paraformaldehyde, washed in cell staining media (CSM) twice, followed by one wash in PBS and one wash in PBS+0.02\% saponin. Blocking was performed with Human TruStain FcX receptor blocking solution (Biolegend, 422302). Cells underwent surface staining with the following surface markers: CD3, CD10, CD19, CD20, CD22, CD24, CD34, CD38, CD43, CD45, CD58, CD61, CD79b, CD123, CD127, CD179a, CD179b, Human Leukocyte Antigen-DR (HLA-DR), Immunoglobulin M (IgM), and Thymic Stromal Lymphopoietin Receptor (TSLPr). 

After surface staining, cells were washed, permeabilized, and intracellular stained with the following markers: Intracellular Immunoglobulin M (IgMi), Marker of Proliferation Ki-67 (Ki67), lambda chain of the B-cell receptor, Paired Box Protein 5 (PAX5), Recombination Activating Gene 1 (RAG1), and Terminal deoxynucleotidyl transferase (TdT). Once intracellularly stained, samples were washed in CSM, Iridium intercalated, washed in CSM, followed by two washes in ultra-pure double-distilled water. To prepare for acquisition, cells were resuspended with normalization beads. Mass cytometry data were then acquired on a Helios (a 3rd generation CyTOF). 

\subsubsection{Data cleaning and preprocessing}

Standard preprocessing steps for mass cytometry data analysis were performed as described previously \cite{cytofmethods2021}. Specifically, ion counts were transformed using the hyperbolic arcsine function with a cofactor of 5, and all markers were scaled to their 99.9th percentile for comparability between markers. Additionally, all cells expressing a marker value over the 99.9th percentile were excluded from analysis in order to remove technical artifacts and outliers. All analyses were performed using the R package tidytof \cite{keyes2023tidytof}.

\subsubsection{Model fitting}

\textbf{Model specification:} MMIL was applied to the ALL cohort via the \emph{Mixture Lasso}, a lasso logistic regression model trained using the algorithm described in \ref{sec:method}. We used a value of ($\rho = 0.75$) for all MMIL models, relying on the clinical knowledge that ALL patients with more than 25\% blasts in their bone marrow receive the clinical diagnosis of leukemia \cite{berg1992leukemiathreshold}. $\zeta$ was estimated using the training set of each fold (see below). 

In addition to Mixture Lasso, two other models were fit and evaluated: the ``optimal'' model (a lasso logistic regression model trained using the true cell labels, as annotated by a pathologist) and the ``naive'' model (a lasso model trained using inherited patient labels instead of cell labels). For clarity, the cell labels used by Mixture Lasso are referred to as the ``probabilistic labels'' of each cell; the cell labels used by the optimal model are referred to as the ``gold-standard'' or ``pathologist-annotated'' labels of each cell; and the labels used by the naive model are referred to as the ``inherited patient labels'' of each cell. Mixture Lasso, the optimal model, and the naive model are referred to as the 3 "model classes" that we evaluated. 

Importantly, for the ALL dataset, only diagnostic bone marrow samples were used to train any models in order to mimic the clinical scenario in which a model is built at the time of diagnosis and used to track a patient's leukemia burden over time.  \newline

\textbf{Hyperparameter tuning:} Lasso models have a single hyperparameter: $\lambda$, the penalty term determining the amount of regularization applied to the model's coefficients. For each model class, we tuned over 10 values of lambda, equally spaced on a logarithmic scale between the lowest value ($10^{-5}$) and the highest value (1). The optimal hyperparameter for each model class was determined using cross-validation (see below). The model predictions of the model fit with the optimal penalty parameter were used for reporting in Figure 4's ROC curves and AUROC values.\newline

\textbf{Cross-validation and model selection:} Model performance was estimated using leave-one-(patient)-out cross-validation (LOOCV), a schema in which all cells from a single ALL patient are held out as a separate test set in each fold of the cross-validation. Specifically, we break the dataset into 3 folds such that each fold includes 2 ALL patients and 3 healthy patients for model training and 1 held-out ALL patient for model evaluation. For each fold and model class, we fit the model on the training set and evaluate its cell-level performance on the held-out ALL patient according to gold-standard cell labels. The optimal lasso regularization penalty for each model class was chosen by selecting the penalty that optimized the average log-likelihood (for Mixture Lasso) or binomial deviance (for both other models, as is ordinarily done) on the held out sample across all folds. Finally, for model interpretation, we refit each model with the optimal penalty on all 3 diagnostic ALL and all 3 healthy bone marrow samples.

\subsubsection{Model evaluation and interpretation}

\textbf{Single-cell model evaluation:} To evaluate each model class's performance at the single-cell level, we calculated the ROC curve and corresponding AUROC for each sample using gold-standard cell labels. 

Although models were only trained using cells from the diagnostic timepoint, the models were evaluated for all available samples at all timepoints. Accordingly, ROC curves and AUROCs from different tissues and timepoints were calculated separately. \newline

\textbf{Model interpretation:} For model interpretation in \textbf{Figure \ref{fig:supplemental_figure_5}}, we examined the nonzero coefficients of the Mixture Lasso model refit on all data using the optimal penalty parameter identified by cross-validation.







\section{Supplementary information}

\subsection{Ethical approval declarations}
This research complies with all relevant ethical regulations. The use of the primary samples was approved by the Institutional Review Board at Lucile Packard Children’s Hospital at Stanford University. Healthy human bone marrow samples (n = 3) were purchased through AllCells. De-identified bone marrow samples from pediatric patients with BCP-ALL were obtained, under informed consent, from the Pediatric Clinic of University of Milano-Bicocca (Centro Maria Letizia Verga, Monza, Italy; n = 3). The use of these primary samples was approved by the Institutional Review Boards at both institutions. Written informed consent was obtained from the parents of the patients or their legal representatives, who agreed to the use of biological material for research and clinical studies. To protect patients’ privacy, samples have been de-identified.

\subsection{Acknowledgments}

EC is supported by the Stanford Data Science Scholars program and a Stanford Graduate Fellowship.
MZ is supported by the Stanford Bio-X Bowes Graduate Student Fellowship.
TH was partially supported by grants DMS-2013736 from the National Science Foundation, and grant R01GM134483 from the National Institutes of Health. 
RT is supported by the National Institutes of Health(5R01EB001988-16) and the National Science Foundation(19DMS1208164).
TJK is supported by the National Institutes of Health [grant number 1F31CA239365-01], the Mark Foundation for Cancer Research, the Andrew McDonough B+ (Be Positive) Foundation, and a Point Foundation Graduate Student Scholarship.
JS is supported by Associazione Italiana per la Ricerca sul Cancro (AIRC; Start-UP grant no. 27325).
GPN is supported by the National Institutes of Health [grant numbers U54CA209971, U54HG010426, and U19AI100627].
KLD is supported by the National Institutes of Health [grant number R01 CA251858-01A1], the Mark Foundation for Cancer Research, the Andrew McDonough B+ (Be Positive) Foundation, the Oxnard Foundation, and the Stanford Maternal and Child Health Research Institute. 

\subsection{Availability of data and materials}

Acute Myeloid Leukemia (AML) mass cytometry data are availability on FlowRespository, with Accession ID FR-FCM-Z2E7.

Acute Lymphoblastic Leukemia (ALL) mass cytometry data are available on Dryad, with DOI 10.5061/dryad.8gtht76vw.

\subsection{Code availability}
An R package to train mixture classifiers can be found on Github in the repo [XXXXXXXX].



\begin{appendices}

\section{Supplemental Figures}

\begin{figure}[!ht]
    \centering
      \includegraphics[width=0.8\columnwidth]{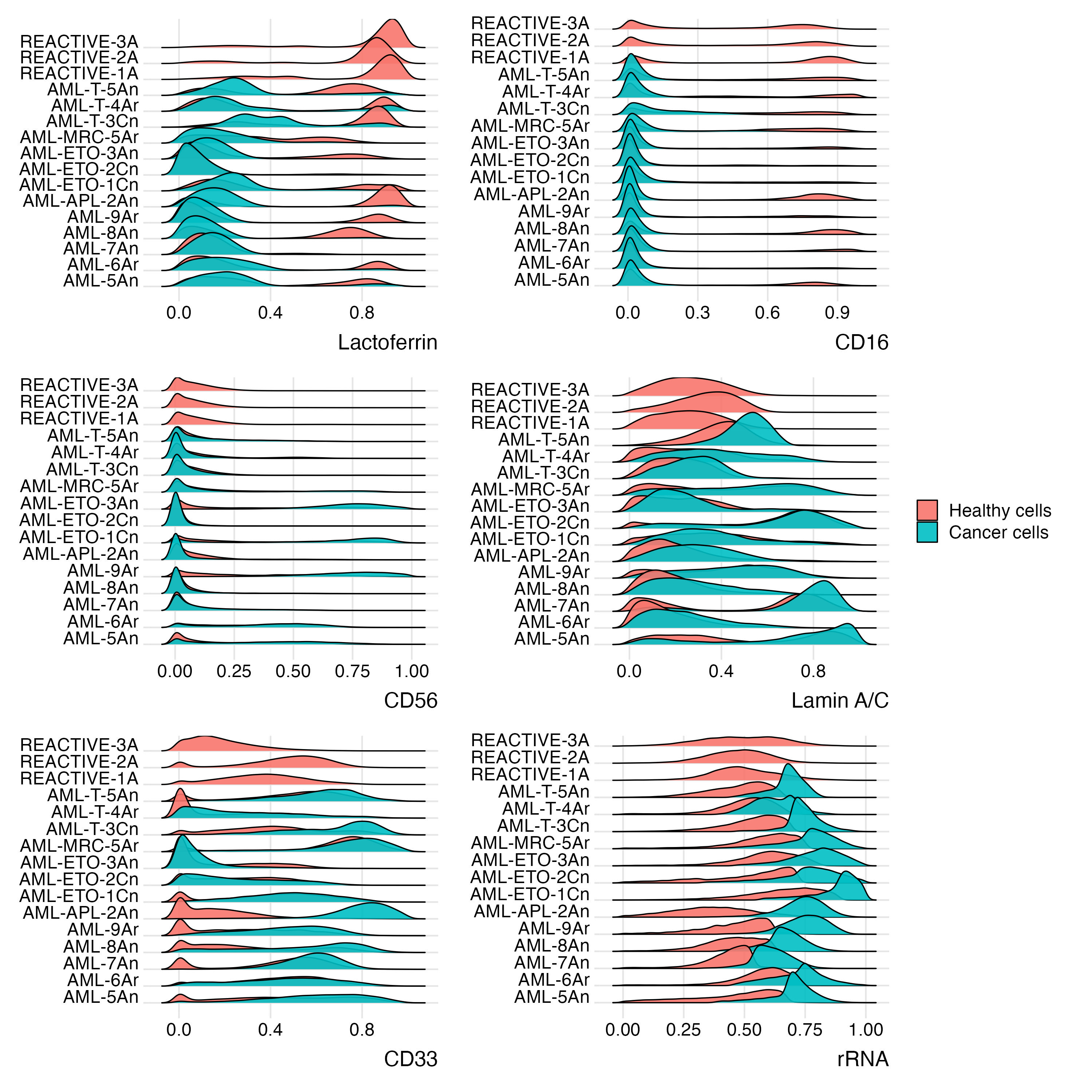}
      \caption{\textbf{Mixture Lasso selects features that discriminate between healthy cells and cancer cells (leukemic blasts) in Acute Myeloid Leukemia (AML).} Density plots indicating that 5 markers (Lactoferrin, CD16, CD56, Lamin A/C, and CD33) selected by Mixture Lasso trained using only patient labels successfully separate healthy and cancer cell populations. Healthy patient IDs are the top 3 density plots in each panel (REACTIVE-3A, REACTIVE-2A, and REACTIVE-3A), and all other plots are AML patients. Note that rRNA, which strongly separates healthy and cancer cells, was not selected by Mixture Lasso, representing an interesting instance of a missed discovery.  
      }
      \label{fig:supplemental_figure_1}
\end{figure}

\newpage

\begin{figure}[!ht]
    \centering
      \includegraphics[width=0.8\columnwidth]{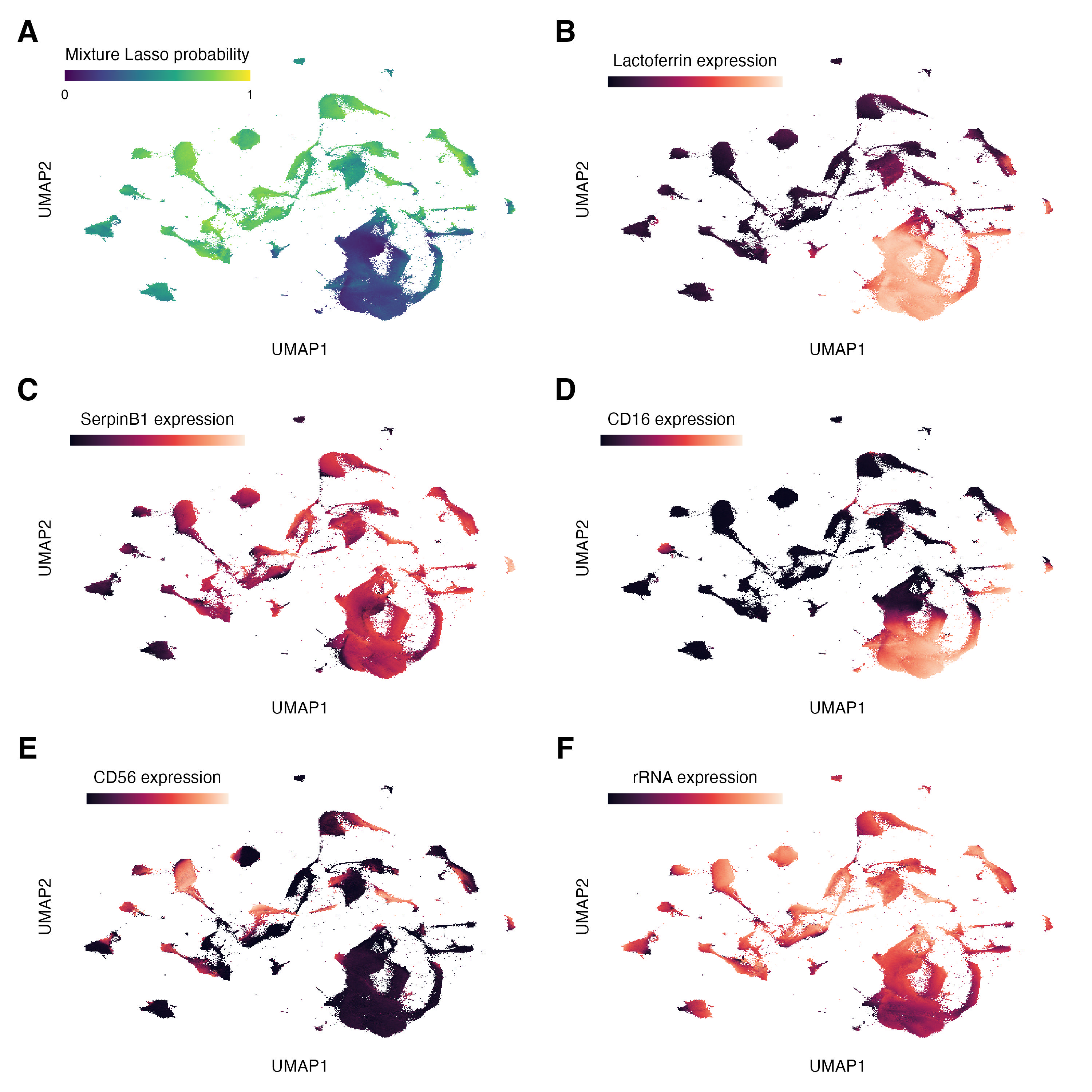}
      \caption{\textbf{Mixture Lasso selects features that distinguish phenotypically distinct cell populations in Acute Myeloid Leukemia (AML).} UMAP diagrams denoting the cell populations expressing 4 of the markers (Lactoferrin, SerpinB1, CD16, and CD56) selected by Mixture Lasso whe trained to detect leukemic blasts in AML using only patient labels. rRNA is also included as one of the most important markers used for manual labeling of leukemic blasts by pathologists in Tsai et al. (2020). Note that, although rRNA was not selected as a disease-associated feature by the Mixture Lasso model, it is strongly expressed by the cells assigned high probabilities by the Mixture Lasso model.}
      \label{fig:supplemental_figure_2}
\end{figure}

\newpage

\begin{figure}[!ht]
    \centering
      \includegraphics[width=0.8\columnwidth]{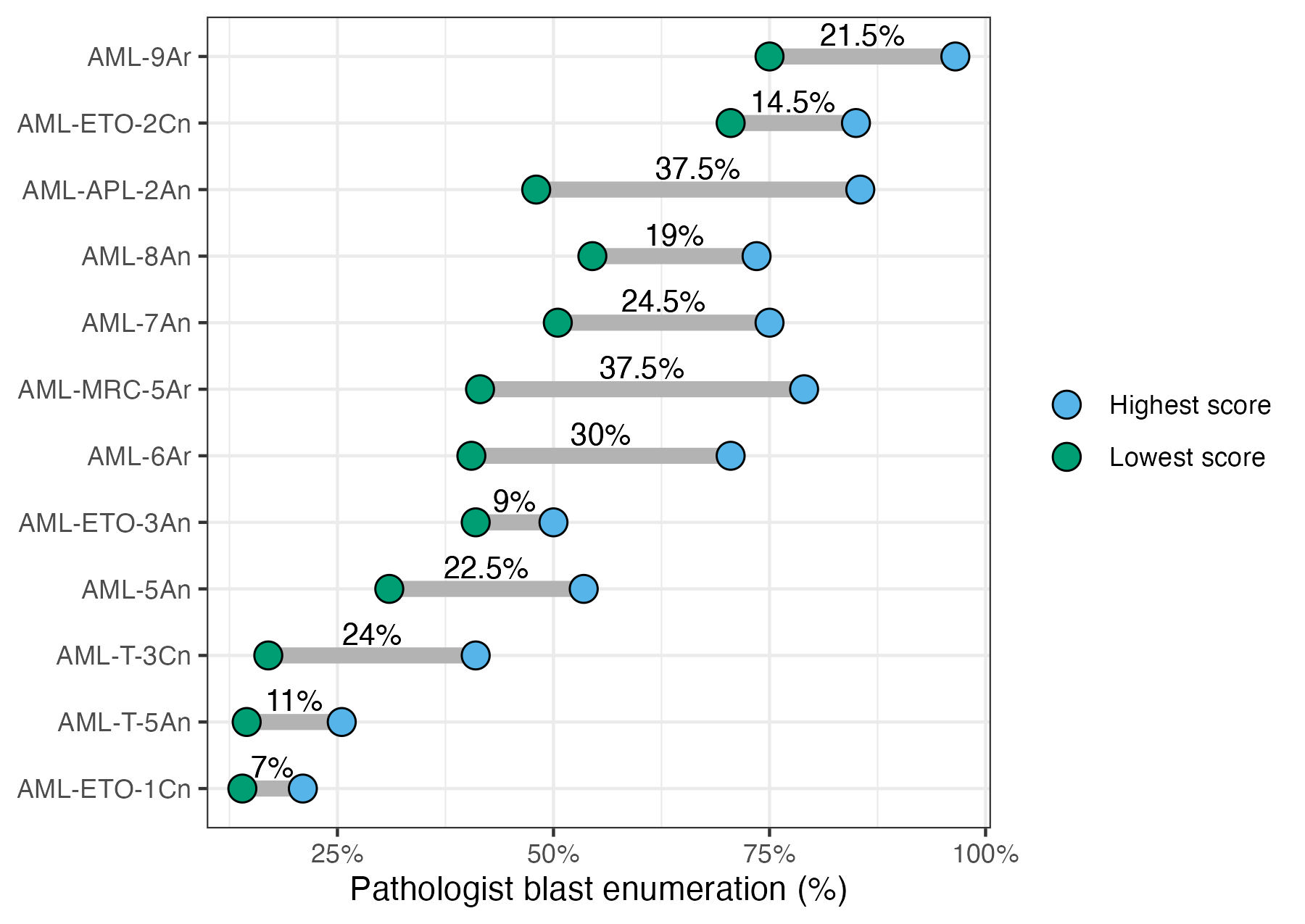}
      \caption{\textbf{Enumeration of leukemic blasts in AML exhibits wide variability among expert pathologists.}
A ball-and-stick plot indicating the range between the highest and lowest blast enumeration scores (percentage of cells) assigned to each AML patient in the Tsai et al. (2020) AML cohort among 4 board-certified hematopathologists. Among these, the median range is 22.5\%.
Blast enumeration was performed using light microscopy, a gold-standard of hematopathological evaluation of leukemia patients. Data were obtained via correspondence with the authorship team of Tsai et al. (2020).}
     \label{fig:supplemental_figure_3}
\end{figure}

\newpage

\begin{figure}[!ht]
    \centering
      \includegraphics[width=0.8\columnwidth]{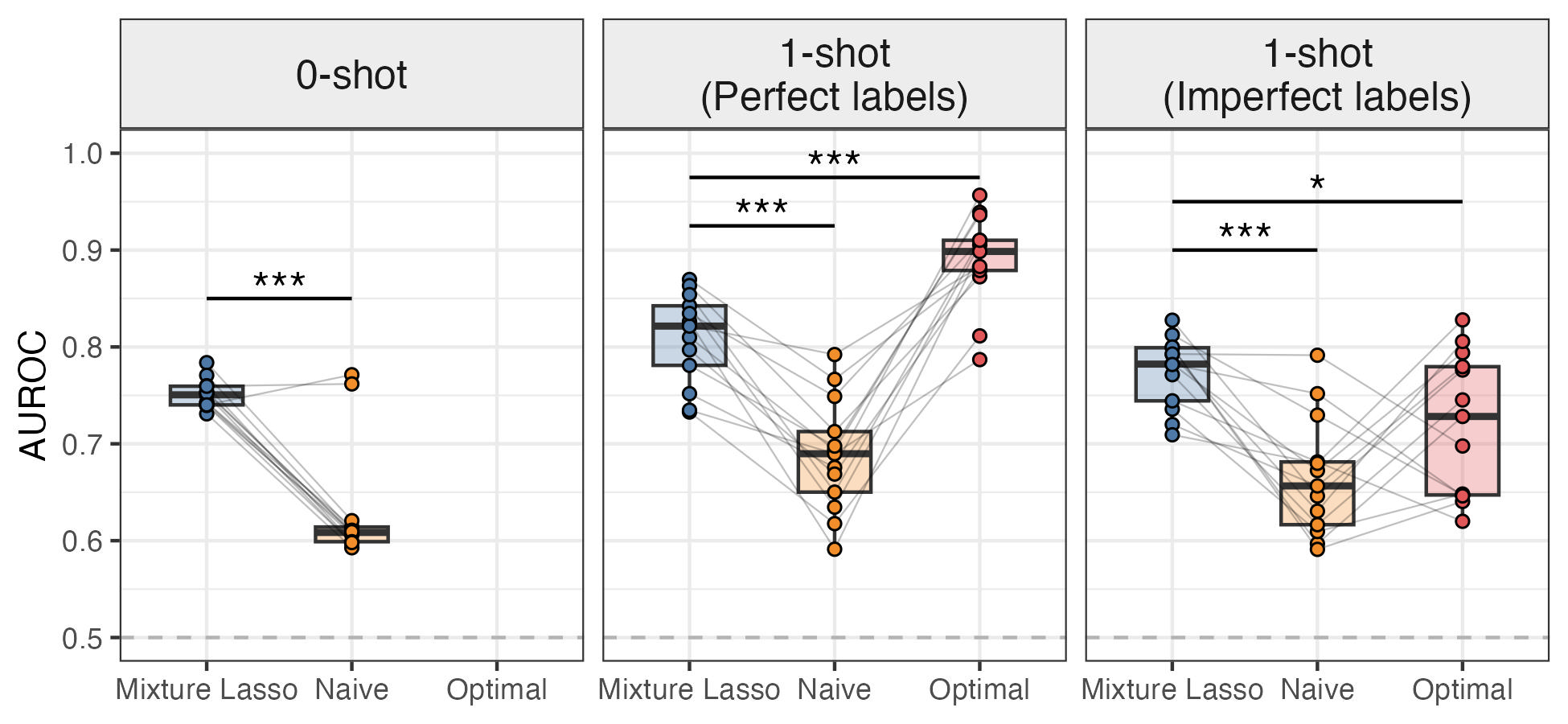}
      \caption{\textbf{1-shot MMIL improves performance at detecting AML blasts by incorporating gold-standard labels while remaining robust to noisy labels.} This plot represents the same result as \ref{fig:figure_3}b, but with AUROC values averaged across patients in the left-out fold (instead of across 1-shot patients). Thus, each point represents the average AUROC across all held-out patients within each 1-shot experiment (rather than the average held-out performance on a given patient across all 1-shot experiments). "*" indicates statistical significance at the level of p < 0.05 and  "***" indicates statistical significance at the level of p < 0.001 using a paired Student's t-test with Benjamini-Hochberg correction for multiple comparisons.
      }\label{fig:supplemental_figure_4}
\end{figure}

\newpage

\begin{figure}[!ht]
    \centering
      \includegraphics[width=0.8\columnwidth]{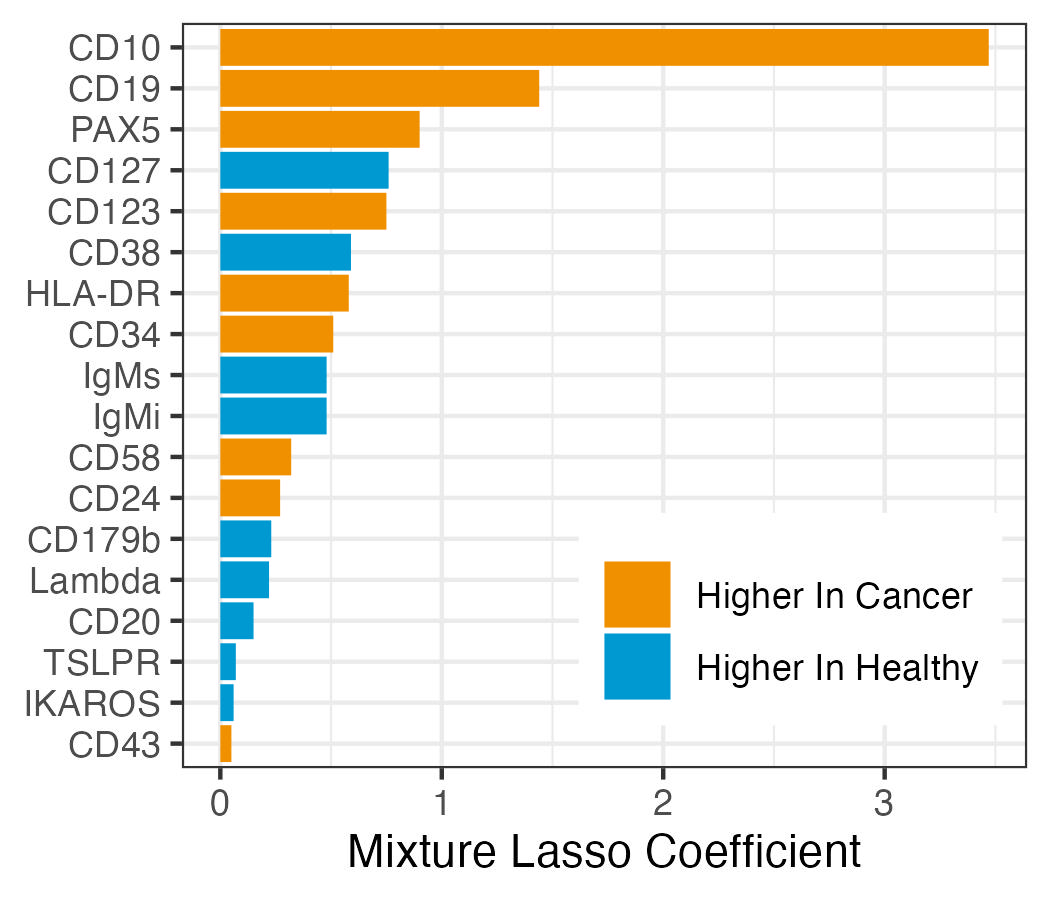}
      \caption{\textbf{Mixture Lasso coefficients in the Acute Lymphoblastic Leukemia model.} 
      }\label{fig:supplemental_figure_5}
\end{figure}

\newpage

\begin{figure}[!ht]
    \centering
      \includegraphics[width=0.8\columnwidth]{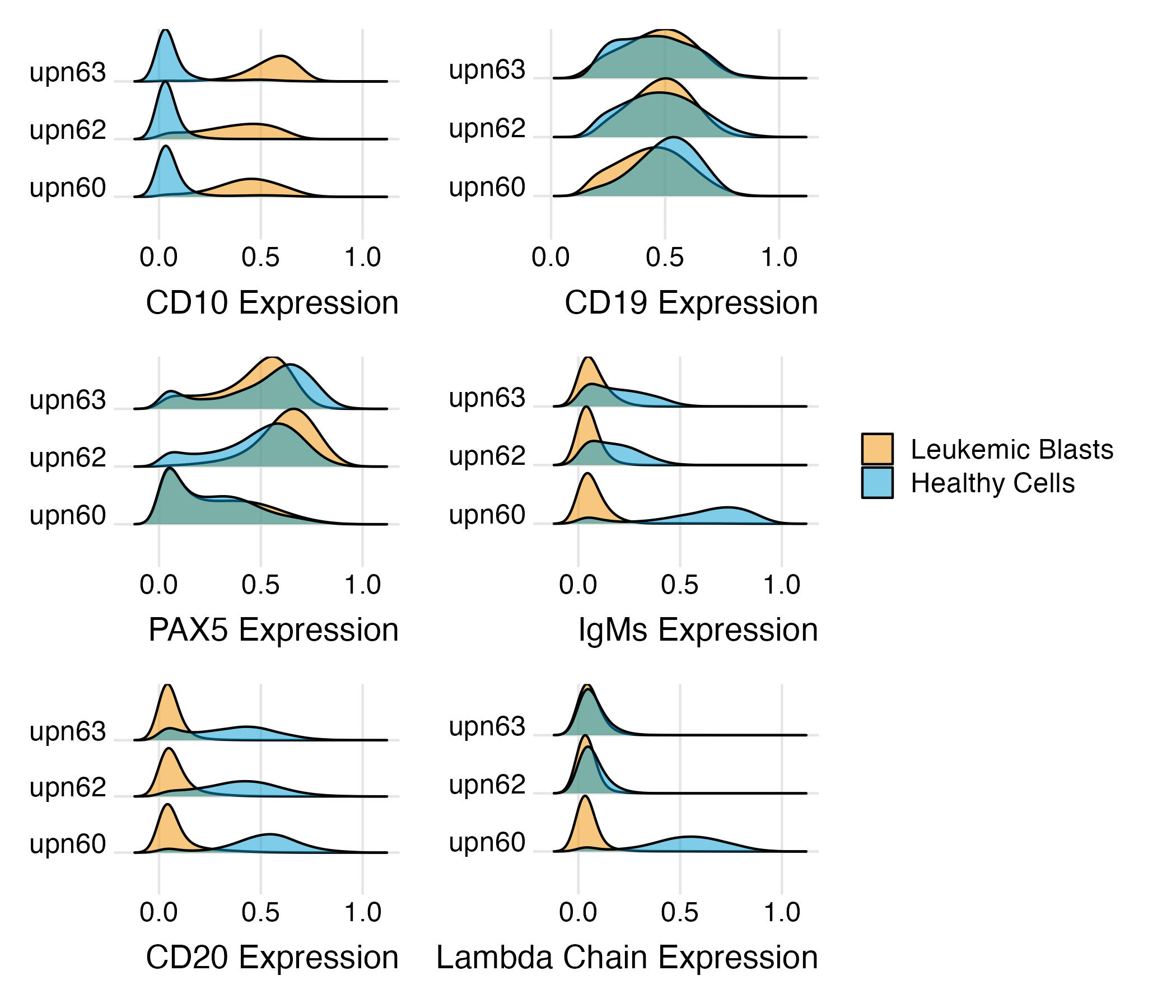}
      \caption{\textbf{Mixture Lasso selects features that discriminate between healthy cells and cancer cells (leukemic blasts) in Acute Lymphoblastic Leukemia (ALL).} Density plots indicating that 6 markers (CD10, CD19, PAX5, IgM (surface), CD20, and Lambda chain (of the BCR receptor) selected by Mixture Lasso trained using only patient labels successfully separate healthy and cancer cell populations across all 3 ALL patients. Patient IDs (upn60, upn62, and upn63) are indicated at left.  
      }\label{fig:supplemental_figure_6}
\end{figure}

\clearpage 

\section{Mixture modeling uses expectation maximization}
\label{sec:likelihoods}
Our approach is an application of expectation maximization~\cite{dempster1977maximum}  to maximize the observed data likelihood, $\prod_{i=1}^n P(z_i \mid x_i)$. The EM procedure is as follows:
\begin{enumerate}
    \item \textit{Expectation step:} At step $t$, use $\eta^{(t-1)}(x)$ to compute $y^{(t)} = P_{\eta^{(t-1)}}(y \mid x, z)$, our current estimates for the unobserved $y$. For cells with $z = 0$, use the known label $y^{(t)} = 0$.
    \item \textit{Maximization step:} Find $\eta^{(t)}$ that maximizes the full likelihood, $\prod_{i=1}^n P(z_i, y_i^{(t)} \mid x_i)$. Adjust $\eta^{(t)}$ to account for biased sampling.
\end{enumerate}

In this section, we will first address the expectation step: we will show how to compute $P(y \mid x, z)$. Then, we will share the details for the maximization step. We will show the intercept adjustment to account for biased sampling, and then use this to write our objective, the observed data likelihood. Finally, we will show that the binomial log likelihood with respect to $y$ is proportional to the full likelihood, and so the maximization step can be achieved by solving a typical binary classification problem.

\subsection{Expectation step: estimating labels}
At the $t^\text{th}$ step of our algorithm, we update our estimate of the unknown labels $y$ as $P_{\eta^{(t-1)}}(y \mid x, z = 1)$, where
\begin{equation}
\label{eq:label_update}
    \text{logit } P_{\eta^{(t-1)}}(y \mid x, z = 1) = \eta^{(t-1)}(x) - \log\left(\frac{\rho \zeta}{1 - (1-\rho)\zeta}\right).
\end{equation}
We note that $\log\left(\frac{\rho \zeta}{1 - (1-\rho)\zeta}\right) < 0$. So, this intercept adjustment raises the baseline probability that a cell is diseased, and this makes sense: knowing a cell came from a sick person \emph{should} make us think the cell is more likely to be disease associated. 

Now, we show our derivation of this label update, and use our assumption that ${P(z \mid y, x) = P(z \mid y)}$. We begin by applying Bayes' rule:
\begin{align}
    P_{\eta^{(t-1)}}(y \mid x, z = 1) &= \frac{P(z = 1 \mid y = 1) P_{\eta^{(t-1)}}(y = 1 \mid x)}{P_{\eta^{(t-1)}}(z = 1 \mid x)} \\
    &= \frac{P_{\eta^{(t-1)}}(y = 1 \mid x)}{P_{\eta^{(t-1)}}(z = 1 \mid x)} \\
    &= \frac{P_{\eta^{(t-1)}}(y = 1 \mid x)}{P(z = 1 \mid y = 1) P_{\eta^{(t-1)}}(y = 1 \mid x) + P(z = 1 \mid y = 0) P_{\eta^{(t-1)}}(y = 0 \mid x)} \\
    &= \frac{P_{\eta^{(t-1)}}(y = 1 \mid x)}{P_{\eta^{(t-1)}}(y = 1 \mid x) + P(z = 1 \mid y = 0) P_{\eta^{(t-1)}}(y = 0 \mid x)}
\end{align}
In the denominator, we used the identity $P(z = 1 \mid y = 1) = 1$. We can compute ${P(z = 1 \mid y = 0)}$:
\begin{equation}
    P(z = 1 \mid y = 0) = \frac{P(y = 0 \mid z = 1) P(z = 1)}{P(y = 0)} = \frac{\rho \zeta}{1 - (1-\rho) \zeta}.
\end{equation}
Now, we can plug this back in to our expression:
\begin{align}
    P_{\eta^{(t-1)}}(y \mid x, z = 1) &= \frac{P_{\eta^{(t-1)}}(y = 1 \mid x)}{P_{\eta^{(t-1)}}(y = 1 \mid x) + \frac{\rho \zeta}{1 - (1-\rho)\zeta} P_{\eta^{(t-1)}}(y = 0 \mid x)} \\
    &= \frac{e^{\eta^{(t-1)}(x)}}{e^{\eta^{(t-1)}(x)} + \frac{\rho \zeta}{1 - (1-\rho)\zeta}}.
\end{align}
With manipulation, this becomes:
\begin{equation}
    \text{logit } P_{\eta^{(t-1)}}(y \mid x, z = 1) = \eta^{(t-1)}(x) - \log\left(\frac{\rho \zeta}{1 - (1-\rho)\zeta}\right).
\end{equation}

\subsection{Maximization step: adjustment for biased sampling}
Relative to the general population, we may have a biased sample of healthy and sick people. To address this, we can update the intercept of our model using the following adjustment:
\begin{equation}
    \label{eq:intercept_adj}
    \text{logit } P(y = 1 \mid x, s = 1) = \eta(x) + \log\left(\frac{(1-\rho) n_1}{n - (1-\rho) n_1}\right) - \log\left(\frac{(1-\rho) \zeta}{1-(1-\rho) \zeta}\right),
\end{equation}
where $s = 1$ indicates our sample, and $n_1$ and $n$ are the number of cells in our sample with label $z = 1$ and the number of cells total, respectively. 

\textit{Proof}: We begin using Bayes rule, and we use the assumption that $P(s \mid y, x) = P(s \mid y)$:
\begin{align}
    P(y = 1 \mid x, s = 1) &= \frac{P(s = 1 \mid y = 1) P(y = 1 \mid x)}{P(s = 1 \mid y = 1) P(y = 1 \mid x) + P(s = 1 \mid y = 0) P(y = 0 \mid x)} \\
    &= \frac{\frac{P(s = 1 \mid y = 1)}{P(s = 1 \mid y = 0)} e^{\eta(x)}}{\frac{P(s = 1 \mid y = 1)}{P(s = 1 \mid y = 0)} e^{\eta(x)} + 1} \\
    &= \frac{e^{\eta^\star(x)}}{e^{\eta^\star(x)} + 1},
\end{align}
where $\eta^\star(x) = \eta(x) + \log\left(\frac{P(s = 1 \mid y = 1)}{P(s = 1 \mid y = 0)}\right)$. Now, we simplify $\log\left(\frac{P(s = 1 \mid y = 1)}{P(s = 1 \mid y = 0)}\right)$:
\begin{align}
    \log\left(\frac{P(s = 1 \mid y = 1)}{P(s = 1 \mid y = 0)}\right) &= \log\left(\frac{\frac{P(y = 1 \mid s = 1) P(s = 1)}{P(y = 1)}}{\frac{P(y = 0 \mid s = 1) P(s = 1)}{P(y = 0)}}\right) \\
    &= \log\left(\frac{P(y = 1 \mid s = 1)}{P(y = 0 \mid s = 1)}\right) - \log\left(\frac{P(y = 1)}{P(y = 0)}\right).
\end{align}
We write out each of these probabilities. We show the full derivation for the first expression; the arguments for the the remainder are similar.
\begin{align}
    P(y = 1 \mid s = 1) &= P(y = 1 \mid z = 1, s = 1) P(z = 1 \mid s = 1) + P(y = 1 \mid z = 0, s = 1) P(z = 0 \mid s = 1) \\
    &= P(y = 1 \mid z = 1) P(z = 1 \mid s = 1) \\
    &=  \frac{(1 - \rho) n_1}{n} \\[.5em]
    P(y = 0 \mid s = 1) &= \frac{n - (1 - \rho) n_1}{n} \\[.5em]
    P(y = 1) &= (1-\rho) \zeta \\[.5em]
    P(y = 0) &= 1 - (1-\rho) \zeta.
\end{align}
Now, we substitute back in:
\begin{align}
    \log\left(\frac{P(y = 1 \mid s = 1)}{P(y = 1 \mid s = 0)}\right) - \log\left(\frac{P(y= 1)}{P(y = 0)} \right)  &= \log\left(\frac{(1-\rho) n_1}{n - (1-\rho)n_1}\right) - \log\left(\frac{(1-\rho) \zeta}{1 - (1-\rho) \zeta} \right).
\end{align}
Therefore, $\eta^\star(x) = \eta(x) + \log\left(\frac{(1-\rho) n_1}{n - (1-\rho)n_1}\right) - \log\left(\frac{(1-\rho) \zeta}{1 - (1-\rho) \zeta} \right)$.

\subsection{Maximization step: the observed likelihood}
The observed likelihood is given by:
\begin{align}
L(\eta \mid z, X) &= \prod_{i=1}^n P(z = z_i \mid x_i, s = 1) \\
    &= \prod_{i=1}^n \left(\frac{e^{\eta^{\star}(x)} + \frac{\rho n_1}{n-n_1(1-\rho)}}{e^{\eta^{\star}(x)} + 1}\right)^{z_i} \left(\frac{1 -  \frac{\rho n_1}{n-n_1(1-\rho)}}{e^{\eta^{\star}(x)} + 1}\right)^{1 - z_i},
\end{align}
where $\eta^{\star}(x) = \text{logit } P(y = 1 \mid x, s = 1) = \eta(x) + \log\left(\frac{(1-\rho) n_1}{n - (1-\rho) n_1}\right) - \log\left(\frac{(1-\rho) \zeta}{1-(1-\rho) \zeta}\right)$, as in Equation~\ref{eq:intercept_adj}.

\textit{Proof: } Consider just one instance, $(x, z)$. We show how to compute ${P(z = 1 \mid x, s = 1)}$, beginning by using the law of total probability and assuming that ${P(z \mid y, s = 1) = P(z \mid y)}$:
\begin{equation}
\label{eq:pz_given_x}
    P(z = 1 \mid x, s=1) = P(z = 1 \mid y = 1) P(y = 1 \mid x, s=1) + P(z = 1 \mid y = 0) P(y = 0 \mid x, s=1) 
\end{equation}
Because all diseased cells come from sick people, $P(z = 1 \mid y = 1) = 1$. To compute ${P(z = 1 \mid y = 0)}$, we will use Bayes' rule. Recall that we assume $P(y = 0 \mid z = 1) = \rho$.
\begin{align}
    P(z = 1 \mid y = 0, s=1) &= \frac{P(y = 0 \mid z = 1) P(z = 1 \mid s = 1)}{P(y = 0 \mid s = 1)} \\
    &= \frac{\rho \frac{n_1}{n}}{1 - (1-\rho)\frac{n_1}{n}} \label{eq:p_y0_z1}
    = \frac{\rho n_1}{n - (1 - \rho) n_1}.
\end{align}
Now, Expression~\ref{eq:pz_given_x} becomes:
\begin{equation}
     P(z = 1 \mid x, s=1) = P(y = 1 \mid x, s=1) +  \frac{\rho n_1}{n - (1 - \rho) n_1} P(y = 0 \mid x, s=1).
\end{equation}
Finally, we use Equation~\ref{eq:intercept_adj}: $P(y = 1 \mid x, s=1) = \frac{e^{\eta^{\star}(x)}}{1 + e^{\eta^{\star}(x)}}$, where ${\eta^{\star} = \eta(x) + \log\left(\frac{(1-\rho) n_1}{n - (1-\rho) n_1}\right) - \log\left(\frac{(1-\rho) \zeta}{1-(1-\rho) \zeta}\right)}$. The contribution of one instance to the likelihood is then:
\begin{align}
     P(z = 1 \mid x, s=1) &= \frac{e^{\eta^{\star}(x)}}{1 + e^{\eta^{\star}(x)}} +  \left(\frac{\rho n_1}{n - (1 - \rho) n_1}\right) \frac{1}{1 + e^{\eta^{\star}(x)}}\\
     &= \frac{e^{\eta^{\star}(x)} + \frac{\rho n_1}{n - (1 - \rho) n_1}}{1 + e^{\eta^{\star}(x)}},
\end{align}
as desired, and $P(z = 0 \mid x, s = 1) = 1 - P(z = 1 \mid x, s = 1)$.

\subsection{Maximization step: the full likelihood}

In the maximization step of the EM, we need to maximize the \emph{full} likelihood using the current expected values of the unknown labels $y$. We will show that the full likelihood $P(y, z \mid X) \propto P(y \mid X)$; this is what allows us to optimize the usual binomial log likelihood at every step of the EM. Suppose we know both $y$ and $z$. Then, the full likelihood is:
\begin{align}
\label{eq:binomial_ll}
L(\eta \mid z, y, X) &= \prod_i P(y = y_i, z = z_i \mid s_i = 1, x_i) \\
    &= \prod_i P(z = z_i, \mid y = y_i, s_i = 1, x_i) P(y = y_i \mid s_i = 1, x_i)  \label{eq:pz_with_y_and_x}\\
    &= \prod_i P(z = z_i, \mid y = y_i, s_i = 1) P(y = y_i \mid s_i = 1, x_i) \label{eq:pz_y_only}\\
    &\propto \prod_i P(y = y_i \mid s_i = 1, x_i). 
\end{align}
To go from Expression~\ref{eq:pz_with_y_and_x} to Expression~\ref{eq:pz_y_only}, we use our assumption that $P(z \mid y, s = 1, x) = P(z \mid y, s = 1)$: there is no systematic difference between cells from healthy and sick people, other than the fact that some cells from sick people are diseased. Now, we show that $P(z = z_i \mid y = y_i, s = 1)$ is a constant, based on our assumptions of the values of $P(z = 1)$ and ${P(y = y_i \mid z = z_i)}$. There are three cases:
\begin{enumerate}
    \item $P(z = 1 \mid y = 1, s = 1) = 1$
    \item $P(z = 1 \mid y = 0, s = 1) = \frac{\rho n_1}{n - n_1(1-\rho)}$ (as in Equation~\ref{eq:p_y0_z1})
    \item $P(z = 0 \mid y = 0, s = 1) = \frac{n - n_1}{n - n_1(1-\rho)}$.
\end{enumerate}
We note that the fourth case, $P(z = 0 \mid y = 1, s = 1)$, is not possible under our assumption that healthy people have no diseased cells. Therefore, to find $\eta$ that maximizes the full likelihood, we only need to find $\eta$ that maximizes the binomial log likelihood.




\section{Modification for learning algorithms that do not allow soft labels} \label{app:no_soft_labels}
For learning algorithms that do not allow ``soft'' labels, do the following. This is as described in the body of this text.

 \begin{enumerate}
     \item Augment your dataset.
     \begin{enumerate}
         \item Make a new covariate matrix $\tilde{X}$ containing one copy of cells from healthy people, and two copies of cells from sick people.
         \item Make a cell label vector $\tilde{y}$ that is $0$ for cells from healthy people. One copy of each cell from sick people should be labeled $1$ and the other $0$.
         \item Make a weight label vector $w^{(0)}$ that is $1$ for cells from healthy people, $\rho$ for cells labeled $0$ from sick people, and $1-\rho$ for cells labeled $1$ from sick people.
     \end{enumerate}
    \item Iterate to convergence. At the $i^{\text{th}}$ step:
    \begin{enumerate}
        \item \emph{Maximization step:} Fit a model $\eta^{\star (i)}(x)$ using $X$, $y$ and $w^{(i-1)}$. Make the intercept adjustment to obtain \[\eta^{(i)}(x) = \eta^{\star (i)}(x) + \log\left(\frac{(1-\rho) n_1}{n - (1-\rho)n_1}\right) - \log\left(\frac{(1-\rho) \zeta}{1 - (1-\rho) \zeta} \right).\]
        \item \emph{Expectation step:} Use $\eta^{(i)}(x)$ to define $w^{(i)}$. Cells from healthy people have $w^{(i)} = 1$. Cell $j$ with label $y = 1$ from a sick person has weight: \[\text{logit } w^{(i)}_j = \eta^{(i)}(x_j) - \log\left(\frac{\rho \zeta}{1 - (1-\rho) \zeta}\right).\] The weight for the corresponding row with label $y = 0$ has weight $1 - w^{(i)}_j$.
    \end{enumerate}
 \end{enumerate}

\end{appendices}

\bibliographystyle{unsrtnat}
\bibliography{bibliography.bib}

\end{document}